\documentclass[12pt]{iopart}
\usepackage{graphicx}
\usepackage{hyperref}
\begin{document}
\title[Time resolution of the plastic scintillator strips with matrix photomultiplier readout
for J-PET tomograph]
{Time resolution of the plastic scintillator strips with matrix photomultiplier readout
for J-PET tomograph}
\author{
P.~Moskal$^1$, O.~Rundel$^1$, D.~Alfs$^1$, T.~Bednarski$^1$, P.~Bia\l as$^1$, E.~Czerwi\'nski$^1$,
A.~Gajos$^1$, K.~Giergiel$^1$, M.~Gorgol$^2$, B.~Jasi\'nska$^2$, D.~Kami\'nska$^1$, \L .~Kap\l on$^{1,3}$,
G.~Korcyl$^1$, P.~Kowalski$^4$, T.~Kozik$^1$, W.~Krzemie\'n$^5$, E.~Kubicz$^1$, Sz.~Nied\'zwiecki$^1$,
M.~Pa\l ka$^1$, L.~Raczy\'nski$^4$, Z.~Rudy$^1$, N.G.~Sharma$^1$, A.~S\l omski$^1$, M.~Silarski$^1$,
A.~Strzelecki$^1$, A.~Wieczorek$^{1,3}$, W.~Wi\'slicki$^4$, P.~Witkowski$^1$, M.~Zieli\'nski$^1$, N.~Zo\'n$^1$}
\address{$^1$ Faculty of Physics, Astronomy and Applied Computer Science, Jagiellonian University, 30-348 Cracow, Poland}
\address{$^2$ Institute of Physics, Maria Curie-Sk\l odowska University, 20-031 Lublin, Poland}
\address{$^3$ Institute of Metallurgy and Materials Science of Polish Academy of Sciences, 30-059 Cracow, Poland.}
\address{$^4$ \'Swierk Computing Centre, National Centre for Nuclear Research, 05-400 Otwock-\'Swierk, Poland}
\address{$^5$ High Energy Physics Division, National Centre for Nuclear Research, 05-400 Otwock-\'Swierk, Poland}
\begin{abstract}
Recent tests of a single module of the Jagiellonian Positron Emission Tomography system (J-PET)
consisting of 30~cm long plastic scintillator strips
have proven its applicability for the detection
of annihilation quanta (0.511 MeV)
with a 
coincidence resolving time (CRT) of 0.266~ns.
The achieved resolution is almost by a factor of two better 
with respect to the current TOF-PET detectors
and it can still be improved since,
 as it is shown in this article,
the intrinsic limit of time resolution 
for the determination of time of the interaction of 0.511~MeV gamma quanta
in plastic scintillators is much lower.
As the major point of the article,
a method 
allowing to record timestamps of several photons,
at two ends of the scintillator strip,
by means of matrix of
silicon photomultipliers 
(SiPM) is introduced. 
As a result 
of
simulations,
conducted with the number of 
SiPM 
varying from 4 to 42,
it is shown that the improvement of timing resolution saturates with the growing number of
photomultipliers, and that the 2~x~5 configuration at two ends allowing to read twenty timestamps,
constitutes an optimal solution.
The  conducted simulations 
accounted for 
the emission time distribution, photon transport and absorption inside the scintillator, 
as well as quantum efficiency and transit time spread of photosensors,
and were checked based on the experimental results.
Application of
the 2~x~5 matrix of SiPM 
allows for achieving the 
coincidence resolving time in positron emission tomography 
of $\approx$~0.170~ns
for 15~cm  axial field-of-view (AFOV)
and $\approx$~0.365~ns
for 100~cm AFOV. 
The results open perspectives for construction of a cost-effective TOF-PET scanner
with significantly better TOF resolution and larger AFOV with respect to the current TOF-PET modalities. 
\end{abstract}
\vspace{2pc}
\submitto{\PMB}
\maketitle

\section{Introduction}

There is a continued interest in improving time resolution of scintillator detectors.
Such improvements are especially challenging in case of the registration of low energy gamma quanta
where the time resolution is limited by the low statistics of scintillation photons.

Superior time resolution for registration of low energy gamma quanta is of crucial importance in the nuclear medicine applications
as e.g. in positron emission tomography (PET), where the new generation of PET scanners utilizes for the image reconstruction
differences between time of flight (TOF) of annihilation quanta from the annihilation 
vertex
to the detectors (
Conti \hyperlink{Conti2009}{2009}, 
Humm et al \hyperlink{Humm2003}{2003}, 
Karp et al \hyperlink{Karp2008}{2008},
Townsend \hyperlink{Townsend2004}{2004}, 
Moses Derenzo \hyperlink{Moses1999}{1999},
Moses \hyperlink{Moses2003}{2003},
Conti Eriksson \hyperlink{ContiEriksson2009}{2009}
).

In order to improve the TOF resolution and to increase a geometrical acceptance of the PET scanners we are developing
a J-PET detection system (
Moskal et al \hyperlink{Moskal2011}{2011} \hyperlink{Moskal2014}{2014} \hyperlink{Moskal2015}{2015},
Raczynski et al \hyperlink{Raczynski2014}{2014} \hyperlink{Raczynski2015}{2015}
).
The system is based on long strips of plastic scintillators which are characterized 
by better timing properties than the inorganic scintillator 
crystals used in the state of the art PET scanners (
Conti \hyperlink{Conti2009}{2009}, 
Humm et al \hyperlink{Humm2003}{2003}, 
Karp et al \hyperlink{Karp2008}{2008},
Townsend \hyperlink{Townsend2004}{2004}, 
)

\begin{figure}[h]
\begin{center}
\includegraphics[width=220pt]{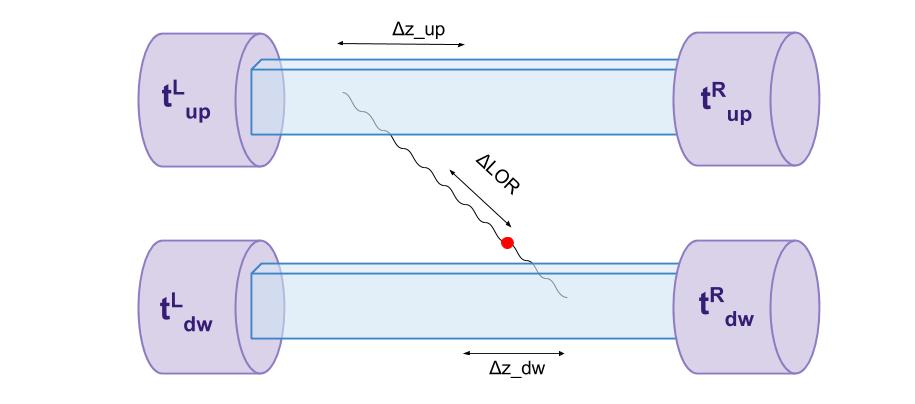}
\includegraphics[width=170pt]{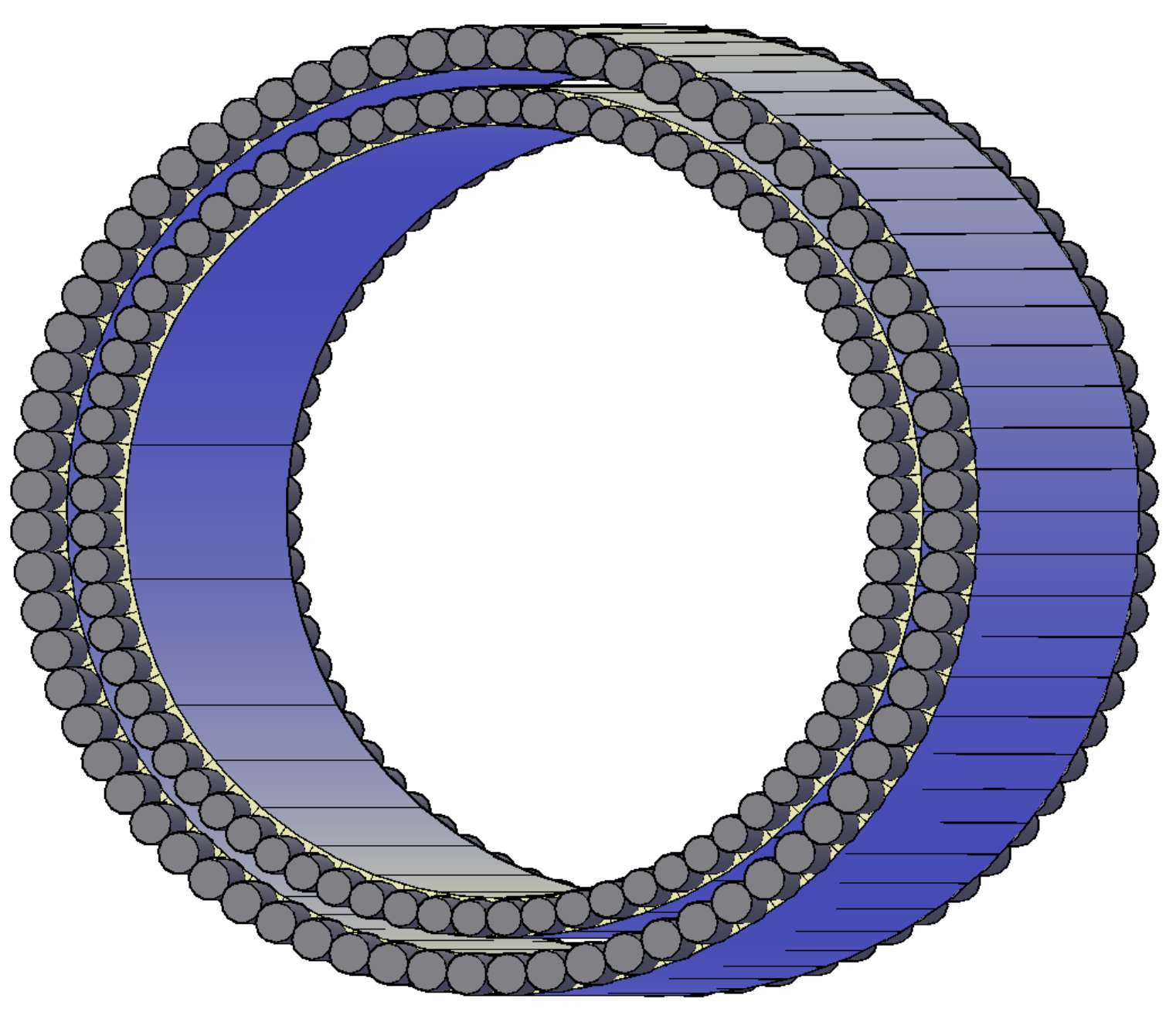}
\end{center}
\caption{
(Left) 
Schematic view of the two detection modules of the J-PET detector.  
A single detection module consists of a scintillator strip read out by two photomultipliers. 
A single event used for the image reconstruction includes information about  
times of light signals arrival to the left (L) and right (R) ends of the upper (up) and lower (dw) 
scintillators (t$^R_{up}$, t$^L_{up}$, t$^R_{dw}$, t$^L_{dw}$). 
A filled red dot inside the figure indicates a place of e$^+$e$^-$ annihilation.  
$\Delta{z}$  denotes the position of the interaction point along the scintillator, 
and $\Delta{LOR}$ indicates the position of annihilation
along the line-of-response (LOR). More details are explained in the text.
(Right) Schematic visualization of an example of two detection layers of the J-PET detector. 
Each scintillator strip is aligned axially and read out at two ends by photomultipliers.
\label{barrel}
}
\end{figure}
Left panel of Fig.~\ref{barrel} shows a schematic view of two detection modules of the \ \hspace{10cm} \ J-PET detector,
where (similarly as described in the 
reference~(Nickles et al \hyperlink{Nickles1978}{1978})) 
the time of the interaction (hit-time) of gamma quantum in the scintillator
($t_{hit} = (t^L + t^R) / 2$)  is calculated as an arithmetic mean of times 
at left ($t^L$) and right ($t^R$) side of the strip.  
The position of interaction along the strip (axial hit position) may be in the first approximation 
calculated  as  $\Delta z$ = ($t^L - t^R$) $v$ / 2, where $v$  denotes the speed of light signals 
in the scintillator strip. For example for plastic strips with cross section of 0.5~cm x 1.9~cm 
the effective light signal velocity is equal to $v$~=~12.2~cm/ns~(Moskal et al \hyperlink{Moskal2014}{2014}).
Thus, in the case of strips with the length of 30~cm characterized with the hit-time resolution 
of  0.188~ns (FWHM) the axial position resolution amounts to  about 2.3~cm (FWHM)~(Moskal et al \hyperlink{Moskal2014}{2014}).
The position along the line-of-response ($\Delta LOR$) between  two strips 
(e.g. up and down shown in the left panel of Fig.~\ref{barrel}) is calculated as 
($\Delta LOR = (t_{hit}^{up} - t_{hit}^{dw}) c / 2$, where $c$ denotes the speed of light.
The hit-time and hence also axial position resolution may still be improved 
e.g. by probing photomuliplier pulses in the voltage domain by a newly developed electronics~(Palka et al \hyperlink{Palka2014}{2014}),
and by applying in the reconstruction the compressing 
sensing theory~(Raczynski et al \hyperlink{Raczynski2014}{2014},\hyperlink{Raczynski2015}{2015})
and the library of synchronized model signals~(Moskal et al \hyperlink{Moskal2015}{2015}).
The timing resolution, as it is introduced hereafter in this article,  
can be also improved by making a readout allowing 
to record time-stamps from larger number of photons compared to the case 
of the vacuum tube photomultipliers.  

Due to the relatively low costs of plastic scintillators and their 
large light attenuation length (in the order of 100~cm) it is possible 
to construct a detector with a long axial field-of-view in a cost effective way.  
This feature makes the J-PET detector competitive to the present solutions as regards the whole-body imaging.
One of the possible arrangements of the scintillator strips in the J-PET scanner
is visualized in the right panel of Fig.~\ref{barrel}.
Such alignment permits to use more than one detection layer thus increasing the efficiency of gamma quanta registration.

\begin{figure}[h]
\begin{center}
\includegraphics[width=250pt]{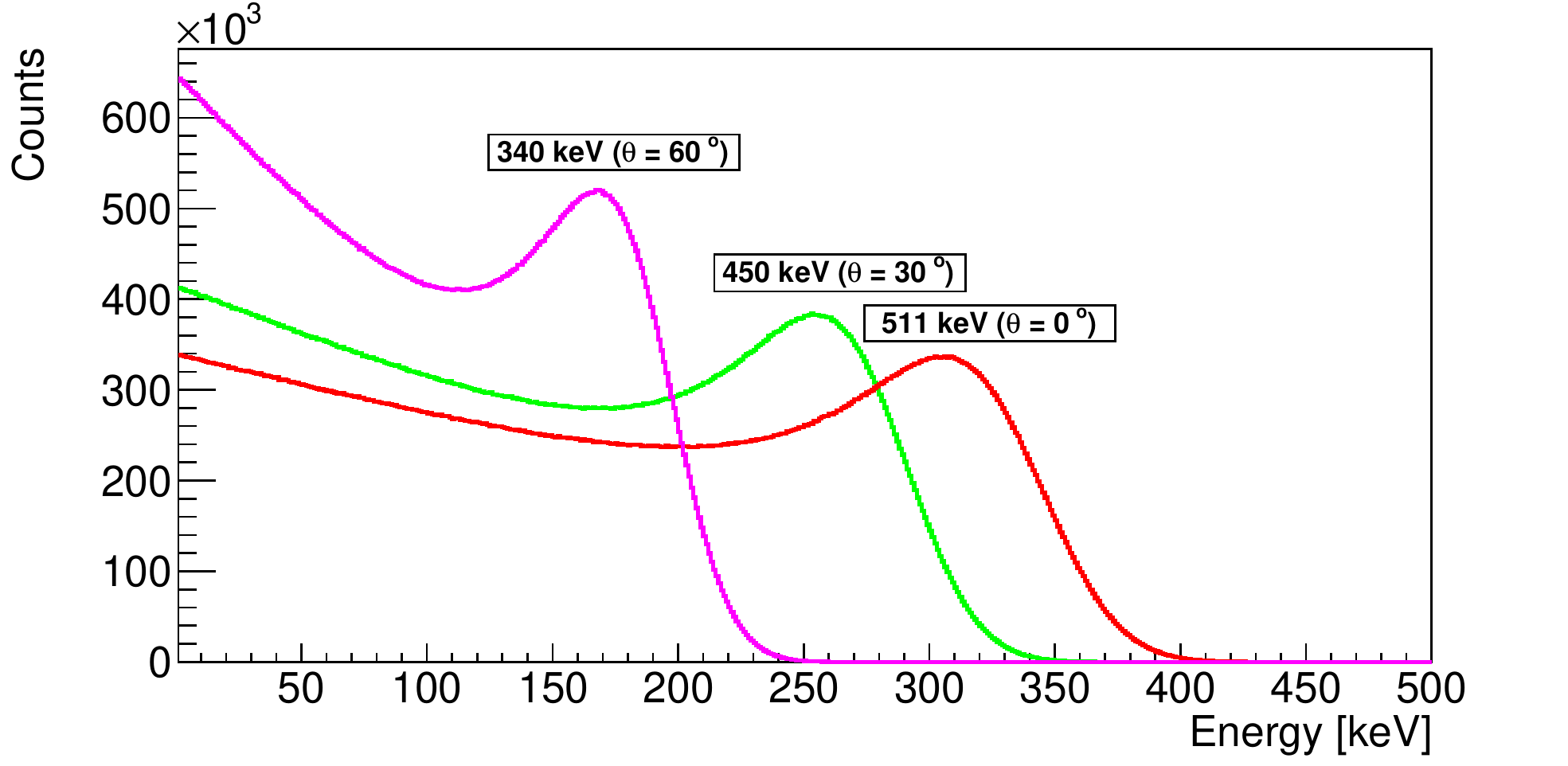}
\end{center}
\caption{
Energy distribution of electrons scattered via the Compton process by gamma quanta 
with energies shown in the plot.  The shown spectra include energy deposition resolution determined 
for the 30~cm long strips of the J-PET detector which is equal 
to $\sigma(E)/E = 0.044 / \sqrt{E(MeV)}$~(Moskal et al {\protect\hyperlink{Moskal2014}{2014}}).
\label{comptons}
}
\end{figure}

Plastic scintillators were not used so for as possible detectors for PET imaging due to the negligible 
probability of the photoelectric effect and lower detection probability with respect to the inorganic crystals.
With the plastic scintillators the detection of 0.511 MeV gamma quanta  is based in practice 
only on the Compton scattering. In Fig.~\ref{comptons} we show a distribution of Compton scattered electron 
energy for (i) the energy of gamma quanta reaching the detector 
without scattering in the patient's body, (ii) after the scattering through 
an angle of 30 degrees and (iii) after scattering through an angle of 60 degrees. 
%
%
The presented distributions show that in order to limit registration of gamma quanta scattered 
in the patient to the range from 0 to about 60 degrees 
(as it was applied earlier e.g. in some LSO or BGO based tomographs~(Humm et al \hyperlink{Humm2003}{2003})),
one has to use an energy threshold of about 0.2 MeV~(Moskal et al \hyperlink{Moskal2012}{2012}).  
The scatter fraction can be further reduced at the expense of the sensitivity, 
yet it should be noted that its suppression to the level achievable 
in the newest LSO based scanners with the energy window of 
0.440-0.625 MeV (Surti et al \hyperlink{Surti2007}{2007}) 
is questionable. 
However, for the quantitative statement, more detailed investigations are needed.
Application of 0.2~MeV threshold suppresses also 
to the negligible level signals due to the secondary 
Compton scattering in the detector material~(Kowalski et al \hyperlink{Kowalski2015}{2015}).
So far we have obtained 0.188~ns ($FWHM$) of hit-time resolution
for the registration of 0.511~MeV annihilation quanta 
by means of 30~cm long strips of plastic scintillators 
read out at both ends by vacuum tube photomultipliers~(Moskal et al \hyperlink{Moskal2014}{2014}).
The experiment was performed using a collimated beam of annihilation quanta 
from the $^{68}Ge$ isotope placed inside a lead collimator with a 1.5~mm wide and 20~cm long slit.  
A dedicated mechanical system allowed us to irradiate the tested  scintillator at the desired position. 
A coincidence registration of signals from the tested module and the reference detector 
on the other side of the collimator ensured selection of annihilation  gamma quanta. 
A tested module consisted of a BC-420 plastic scintillator~(\hyperlink{SaintGobain}{Saint Gobain Crystals})
with dimensions 
of 0.5~cm x 1.9~cm x ~30~cm connected optically at the ends to the Hamamatsu photomultipliers 
R5230~(\hyperlink{Hamamatsu}{Hamamatsu}). 
The experimental setup and results of the measurements are 
described in detail in reference~(Moskal et al \hyperlink{Moskal2014}{2014}).

\begin{figure}[h]
\begin{center}
\includegraphics[width=220pt]{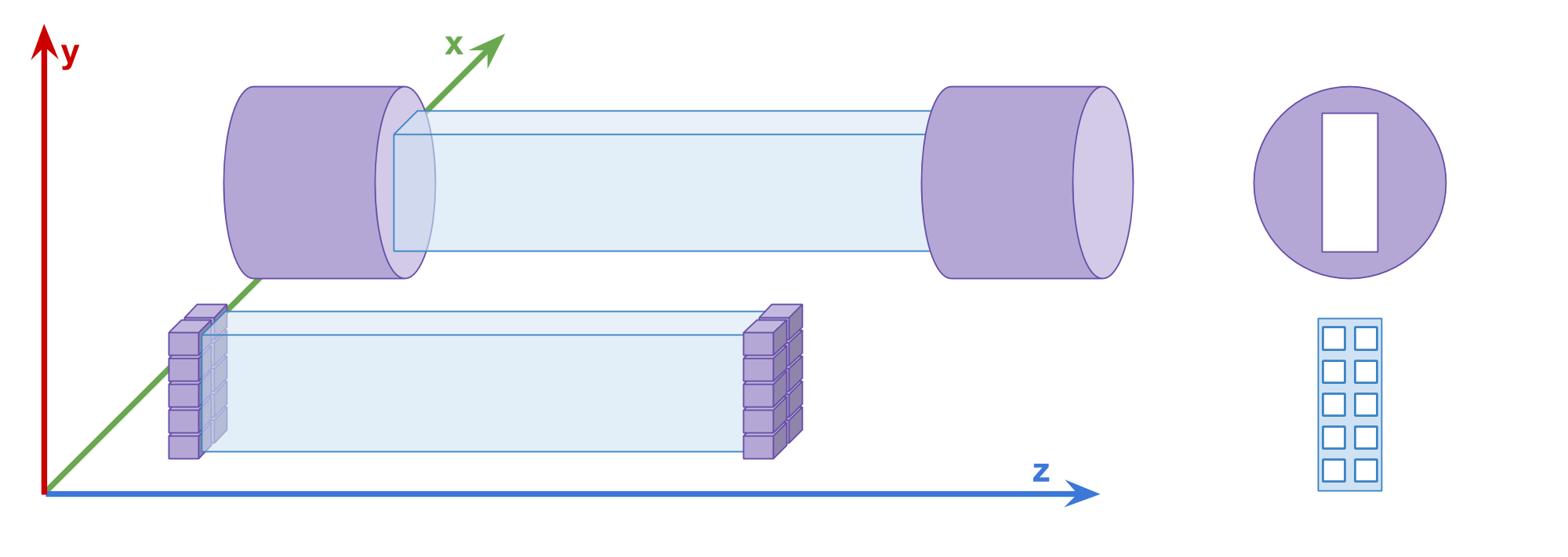}
\end{center}
\caption{
Upper scheme indicates a single module of the J-PET detector consisting of the scintillator 
strip read out on two sides by vacuum tube photomultipliers~(Moskal et al \protect\hyperlink{Moskal2014}{2014}).
Lower part of the figure indicates a scheme of an exemplary multi-SiPM readout allowing for 
determination of timestamps of 20 detected photons (ten on each side). 
Right panel of the figure shows arrangement of photomultipliers. 
Geometrical overlap between the scintillator and the photosensitive 
part of the photomultipliers is marked as white rectangles.
\label{detector}
}
\end{figure}

In principle, information about a time of interaction of gamma quantum in the scintillator
is carried by all 
emitted
scintillation photons.
However, in practice in the typical detectors, only few first registered photons, contributing
to the leading edge of the electrical signal generated by the photoelectric converters,
are utilized in the determination of the onset of these signals
and hence in the determination of the time of the gamma quantum interaction.
This is also the case for the scintillator strips in the current version 
of the J-PET detector (see upper part of Fig.~\ref{detector}),
where the time of the interaction is determined as an arithmetic mean of times at which electric signals generated by
photomultipliers attached to both ends cross a preset threshold voltage.
Therefore, the time resolution may be improved
by making a readout allowing to record timestamps from larger number of photons arriving at the scintillator edge.
There are first attempts to register all timestamps using arrays of single-photon avalanche diodes 
(Meijlink et al \hyperlink{Meijlink2011}{2011}), 
but presently the registration of arrival time of all photons with a good time resolution 
at large areas is still rather impractical.
It is, however, important to stress that
the intrinsic timing resolution limit is approached already
when using only about 20 timestamps from first detected photons
(Seifert et al \hyperlink{Seifert2012}{2012}).
In this article we study
the possibility of improving the time resolution for the large size detectors (few tens of centimeters) 
by registration of timestamps 
from several photons.
This may be realized by preparing a readout in the form of an array with several SiPM 
photomultipliers as it is indicated in the lower part of Fig~\ref{detector}. 
In such a case,
a set of all registered photons is divided into several subgroups and
a time of the registration of the first photon in each subgroup is recorded.
This allows to construct estimators of the time of gamma quantum interaction
based on the number of timestamps equal to the number of SiPM photomultipliers.

In the following sections, 
first we estimate time resolution limits
for infinitesimally small detector making an idealistic assumption that
the time of arrival of each photon can be measured
and used for the estimation of the time of gamma quantum interaction.
We use Fisher information
from all emitted photons and calculate the Cram\'e{}r-Rao lower limit (
Seifert et al \hyperlink{Seifert2012}{2012},
De~Grot \hyperlink{DeGroot1986}{1986}
)
which is independent of the estimator used for the time resolution determination. 
Such estimations of the lower bound for the time resolution 
have been published recently (Seifert et al \hyperlink{Seifert2012}{2012}) for small size crystals, 
taking into account the transit time spread of photomultipliers 
and neglecting the spread of the transit time inside scintillators. 
In this article we extend these investigations to the plastic scintillators strips with the length of up to 100~cm
and include in the estimations the transit time spread due to the propagation 
of photons in scintillator strips as well as the transit time spread of photomultipliers.
Further on we determine the lower bounds for the time resolution using a weighted mean of timestamps
as an estimator of the time of the gamma quantum interaction. 
Next, we describe parameters used in the realistic simulations,
including time distribution of photons emitted in ternary plastic scintillators,
losses and time spread of photons due to their propagation through the scintillator, as well as
quantum efficiency and transit time spread of photomultipliers. We test the simulation procedures
by comparing simulated and experimental results for the BC-420 plastic scintillator read out at two ends 
by Hamamatsu 
R5320
photomultipliers. 
Section \ref{mainsection} contains description of the main idea of this article
where the estimator of the time of the gamma quantum interaction
is defined based on the time ordered set including timestamps from first photons registered
by the matrix of SiPM converters. In this section we perform realistic simulations for the BC-420 scintillator strip
and various configurations for the arrays of S12572-100P Hamamatsu silicon photomultipliers.
Finally we estimate time resolution as a function of the scintillator length 
for the multi-SiPM readout allowing to determine timestamps of 20 photons. The results are compared to the resolutions
achievable with the traditional readout with the vacuum tube photomultipliers.  

The light yield of plastic scintillators amounts to about 10000 photons per 1~MeV of deposited energy.
Annihilation gamma quanta (0.511~MeV) used for the positron emission tomography 
interact with plastic scintillators predominantly via Compton scattering 
(Szymanski et al \hyperlink{Szymanski2014}{2014}),
and therefore may deposit maximally an energy of 0.341~MeV (2/3 of electron mass).
This corresponds to the emission of about 3410 photons. On the other hand, in order 
to decrease the noise 
due to the scattering of gamma quanta inside a patient body a minimum energy deposition 
of about 0.2~MeV is required (Moskal et al \hyperlink{Moskal2012}{2012}). 
Therefore, number of emitted photons discussed hereafter 
in this article includes the range from 2000 to 3410 photons.
                                                                                                     
\section{Estimator of hit-time resolution (variance)}
A single detection module considered in this article consists of the plastic scintillator strip
connected at two ends to photomultipliers (see Fig.~\ref{detector}). 
We assume that the gamma quantum 
or any other particle of interest interacts in the scintillator 
at time~$\Theta$. 
We consider the
resolution for the reconstruction of the value of $\Theta$ based on time measurement of signals generated 
by photosensors attached to two scintillator ends.
For practical reasons, if applicable, we use notation analogous 
to the one introduced in references (Seifert et al \hyperlink{Seifert2012}{2012}).

In general, timestamps of all photons detected at the left ($t^L_1, t^L_2, ..., t^L_{N_L}$) 
and at the right side ($t^R_1, t^R_2, ..., t^R_{N_R}$) 
of the scintillator may be used
for the estimation of the time of the interaction $\Theta$.
It is advantageous to order the sets of timestamps according to ascending time such that: 
($t^L_{(1)} \le t^L_{(2)} \le ... \le t^L_{(N_L)}$)  and 
($t^R_{(1)} \le t^R_{(2)} \le ... \le t^R_{(N_R)}$),
where indices in brackets indicate timestamps from the ordered set.
The $t_{(i)}$ element in the ordered set is referred to as i-th order statistic
(Seifert et al \hyperlink{Seifert2012}{2012}). 
After ordering, the timestamps on one side become correlated  
but the ordered set allows for the simple and intuitive estimation
of time difference between the signal arrivals to the ends of the scintillator:
\begin{equation}
   \Delta t_{(i)} = t^L_{(i)} - t^R_{(i)},
\label{delta_t_i}
\end{equation}
as well as  interaction time $\Theta$ which may be estimated by:
\begin{equation}
   \Theta_{(i)} = \frac{t^L_{(i)}+t^R_{(i)}}{2} + const_{(i)},
\label{Theta_i}
\end{equation}
where $const_{(i)}$ is subject to calibration and for simplicity,
but without loss of generality, it will be omitted in the further considerations.
Distributions of ordered timestamps at one side (e.g. $t^L_{(i)}$ and $t^L_{(j)}$ for $i\ne j$)
are correlated and not identical. However,  
distributions for the same order statistics at left and right side 
($t^L_{(i)}$ and $t^R_{(i)}$) 
are uncorrelated 
since the ordering at left side was done independently of the ordering at right side.
Hence, it follows that variances of  
$\Delta t_{(i)}$, and
$\Theta_{(i)}$ may be expressed as: 
\begin{equation} 
\sigma^2(\Delta t_{(i)}) = \sigma^2(t^L_{(i)}) + \sigma^2(t^R_{(i)}) -2cov(t^L_{(i)},t^R_{(i)}) = \sigma^2(t^L_{(i)}) + \sigma^2(t^R_{(i)}),  
\label{variances1}
\end{equation} 
\begin{equation} 
\sigma^2(\Theta_{(i)}) = \frac{1}{4} 
\left( \sigma^2(t^L_{(i)}) + \sigma^2(t^R_{(i)}) + 2cov(t^L_{(i)},t^R_{(i)}) \right) = 
\frac{1}{4} \left( \sigma^2(t^L_{(i)}) + \sigma^2(t^R_{(i)}) \right).  
\label{variances2}
\end{equation} 
The above equations imply that:
\begin{equation} 
\sigma^2(\Delta t_{(i)}) = 4 \sigma^2(\Theta_{(i)}) = \sigma^2(t^L_{(i)}) + \sigma^2(t^R_{(i)})
\label{variances3}
\end{equation} 
We have checked this supposition by numerical simulations for the probability density 
distributions of emission times considered in this article.
Therefore, as regards the variance, it is sufficient to study properties of only one of these estimators.
Moreover, in order to facilitate direct comparison with results published in the field of TOF-PET
we will express resolution as FWHM of coincidence resolving time (CRT), 
where 
coincidence resolving time determined for i-th order statistic
will be referred to as 
CRT$_{(i)}$.
It should be, however, noted that in general, even though 
$t^L_{(i)}$ and $t^R_{(i)}$ are uncorrelated,
the
$\Delta t_{(i)}$ and  
$\Theta_{(i)}$  may be correlated since
cov($\Delta t_{(i)}$,$\Theta_{(i)}$) =  ($\sigma^2(t^L_{(i)})$ - $\sigma^2(t^R_{(i)})$)/2
is equal to zero only if the emission point is in the center of the detector 
because only in this case 
the 
$t^L_{(i)}$ and $t^R_{(i)}$ are identically distributed. 

In next sections we define the emission time distribution for the plastic scintillators and estimate 
the
Cram\'e{}r-Rao 
lower limit for the achievable time resolution.
Further on we will simulate time resolution for each order statistics 
$\Delta t_{(i)}$ separately, and we will test the variance of the weighted mean of 
$\Delta t_{(i)}$ values showing that such estimator of 
$\Delta t$ allows to reach significantly better resolution than achievable with single order statistics.

\section{Emission time distributions}
\begin{figure}
\begin{center}
\includegraphics[width=220pt]{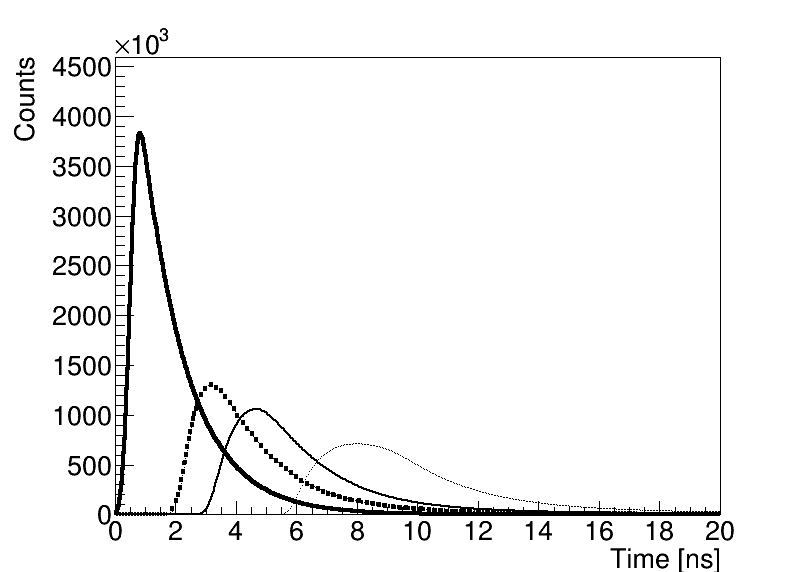}
\end{center}
\caption{
Thick solid line denotes the time distribution for photons simulated according to the formula~\ref{emissiontime} 
describing the probability density distribution for ternary plastic scintillators 
with parameters adjusted to the properties of BC-420 scintillator.
Thick dashed line indicates distribution at 30~cm from the interaction point,
thin solid at 50~cm, and
thin dashed at 100~cm.
These probability density distributions  were simulated taking into account the time of photons propagation through 
a given distance along the scintillator with cross section of 0.5~cm x 1.9~cm. 
Simulations are described in greater details in the Appendix.
\label{emissiontimedistribution}
}
\end{figure}

In case of the ternary plastic scintillators, as e.g. BC-420
(\hyperlink{SaintGobain}{Saint Gobain Crystals})
  and its equivalent EJ-230
(\hyperlink{Eljen}{Eljen Technology})
used in the J-PET detector, 
the distribution of the time of the 
photon emission 
followed by the interaction of the gamma quantum at time~$\Theta$
can be well approximated by the following convolution of gaussian and exponential terms
(Moszynski Bengtson \hyperlink{Moszynski1977}{1977} \hyperlink{Moszynski1979}{1979}):
\begin{equation}
    f(t|\Theta) = K \int_\Theta^t{(e^{-\frac{t-\tau}{t_{d}}}-e^{-\frac{t-\tau}{t_r}}) 
            \cdot e^{-\frac{(\tau-\Theta-2.5\sigma)^2}{2\sigma^2}} d\tau},
\label{emissiontime}
\end{equation}
where the gaussian term with the standard deviation~$\sigma$ reflects the rate of energy transfer 
to the primary solute, whereas $t_r$ and $t_d$ denote the average time of the energy transfer to the
wavelength shifter, and decay time of the final light emission, respectively
(Moszynski Bengtson \hyperlink{Moszynski1979}{1979}). 
K stands for the normalization constant  ensuring that 
$\int_{\Theta}^{+\infty}f(t|\Theta) dt = 1$.
We have set 
$t_d = 1.5$~ns and 
treated $t_r$ and $\sigma$ as a phenomenological  parameters
and adjusted their values to: 
$t_r = 0.005$~ns, 
$\sigma = 0.2$~ns 
in order to describe the properties of the light pulses from the BC-420 scintillator 
i.e. rise time of 0.5~ns, decay time $t_d~=~1.5$~ns and FWHM of 1.3~ns
(\hyperlink{SaintGobain}{Saint Gobain Crystals}).
The resultant distribution of the emission time is indicated by the black solid
line in Fig.~\ref{emissiontimedistribution}. 
Other lines in this figure correspond to time distributions of photons 
in the scintillator with the cross section of 0.5~cm x 1.9~cm 
simulated at 
various distances 
from the interaction point. 
These distributions will be used in the next section for the estimation of the 
lower limits of the achievable time resolutions. 

\section{Cram\'e{}r-Rao lower limit on the resolution of hit-time reconstruction}
The time resolution achievable with the scintillator detectors is limited by the 
optical and electronic time spread caused by the detector components, 
and by the time distribution of photons contributing to the formation of electric signals.
The latter depends on the number of registered photons and 
is referred to as the photon counting statistics
(Seifert et al \hyperlink{Seifert2012}{2012}).
Limitations of the time resolution due to the photon counting statistics have been studied
in detail e.g. in refs.(
Seifert et al \hyperlink{Seifert2012}{2012},\hyperlink{Seifert2012b}{2012b)},
Spanoudaki Levin \hyperlink{Spanoudaki2011}{2011},
Fishburn et al \hyperlink{Fishburn2010}{2010}
)
and the comprehensive account on this topic may be examined e.g. in ref.(Seifert et al \hyperlink{Seifert2012}{2012}).
To large extent this research is driven by the endeavor to improve the 
timing properties of the PET systems
(
Conti Eriksson \hyperlink{ContiEriksson2009}{2009},
Moszynski et al \hyperlink{Moszynski2011}{2011},
Schaart et al \hyperlink{Schaart2009}{2009},\hyperlink{Schaart2010}{2010},
Kuhn et al \hyperlink{Kuhn2006}{2006},
Lecoq et al \hyperlink{Lecoq2013}{2013}), 
and therefore so far
the investigations concentrated on the small size crystal scintillators. In the recent work 
a detailed elaboration of the lower bound for time resolution
has been published for most kinds of available crystal scintillators
(Seifert et al \hyperlink{Seifert2012}{2012}). 
The estimation included transit time spread of photomultipliers
but  
the spread due to the 
transport of photons inside the scintillators was neglected. This was justified since only small size crystals  
(in the order of 1~cm or smaller) were considered.
Here, inspired by the new solution for the PET system based on plastic scintillators 
(Moskal et al \hyperlink{Moskal2011}{2011},\hyperlink{Moskal2014}{2014},\hyperlink{Moskal2015}{2015},
Raczynski et al \hyperlink{Raczynski2014}{2014},\hyperlink{Raczynski2015}{2015}),
we extend the studies of Seifert and coauthors (Seifert et al \hyperlink{Seifert2012}{2012}) from the small size crystals 
to the large size plastic scintillators.
In this section we estimate the lower limit of the time resolution  achievable with scintillator strips of up to 100~cm
assuming ideal electronic systems, and further on 
in the following sections we describe 
results of realistic simulations 
conducted taking into account both photon transport in scintillator material and transit time spread in photomultipliers.

The variance of any unbiased estimator $\Xi$ of the hit-time $\Theta$ 
satisfies the Cram\'e{}r-Rao inequality
(De~Groot \hyperlink{DeGroot1986}{1986},
Seifert et al \hyperlink{Seifert2012}{2012}):
\begin{equation}
    var(\Xi) \ge \frac{1}{I_N(\Theta)},
\label{cramer-rao}
\end{equation}
where $I_N(\Theta)$ denotes the Fisher information concerning $\Theta$ in the set of N randomly chosen timestamps. 
This very general formula enables calculation of the lower bound of the variance of unbiased estimator. 
In case of point estimation of a parameter it quantitatively informs about the estimation efficiency
and whether there is room for improvement.
Knowing the probability density distribution of the photon registration time t following the gamma quantum interaction 
time $\Theta$: $f(t|\Theta)$, the Fisher information in the sample of 
N independent timestamps reads
(De~Groot \hyperlink{DeGroot1986}{1986},
Seifert et al \hyperlink{Seifert2012}{2012}):
\begin{equation}
I_N(\Theta)=N \int_{-\infty}^{+\infty}\frac{(\frac{\partial}{\partial\Theta}f(t|\Theta))^2}{f(t|\Theta)}dt
\label{fisherinfo}
\end{equation}

\begin{figure}
\begin{center}
\includegraphics[width=220pt]{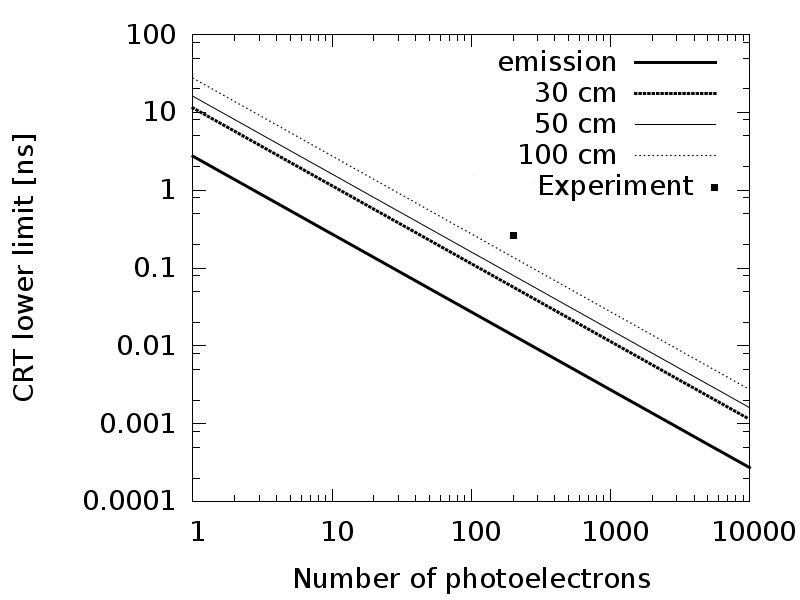}
\end{center}
\caption{ 
Cram\'e{}r-Rao lower limit 
for the time resolution achievable with plastic scintillators calculated as a function of 
number of registered photons and as a function of the scintillator length assuming cross section of 0.5~cm~x~1.9~cm. 
The meaning of the curves is described in the legend.  
The square indicates time resolution determined experimentally using a first version of the J-PET prototype
with plastic scintillator strips of dimensions 
0.5~cm~x~1.9~cm~x~30~cm~(Moskal et al \protect\hyperlink{Moskal2014}{2014}).
The result does not include the time spread due to the unknown depth-of-interaction.
\label{figlimitteoria}
}
\end{figure}

Fig.~\ref{figlimitteoria} shows the lower limit of the time resolution estimated as a function of the number of registered photons N 
based on relations~\ref{cramer-rao}~and~\ref{fisherinfo}. 
Thick-solid line shows results assuming that the time distribution of registered photons is the same 
as the emission time distribution indicated by the thick solid line in Fig.~\ref{emissiontimedistribution}.
The other lower limits shown with thick-dashed, 
thin-solid and thin-dashed lines were obtained assuming 
time distributions of photons after passing a distance of 30~cm, 
50~cm and 100~cm of the plastic scintillator strip with a cross section of 0.5~cm~x~1.9~cm. 
The corresponding time distributions are shown 
in Fig.~\ref{emissiontimedistribution}.    

In the limit of only one photon the result is quite intuitive since in this case the 
Cram\'e{}r-Rao lower bound
corresponds to about 3~ns which is approximately in the order of 
the FWHM of the time distribution of the emitted photons 
(solid line in Fig.~\ref{emissiontimedistribution}) 
amounting to $\sim$~3.5~ns.
The superimposed square indicates an experimental result obtained for the strip of BC-420 plastic 
scintillator with the dimensions 
of 0.5~cm x 1.9~cm x 30~cm read out by Hamamatsu 
R5320
photomultipliers~(Moskal et al \hyperlink{Moskal2014}{2014}).  
A comparison with the corresponding lower limit implies that there is still room for the substantial improvement 
of the time resolution.  
In the next sections we will present a novel solution which allows to improve the time resolution 
by more than a factor of 1.5.

\section{Time resolution as a function of the order statistics for the ideal plastic scintillator}
\begin{figure}
\begin{center}
\includegraphics[width=220pt]{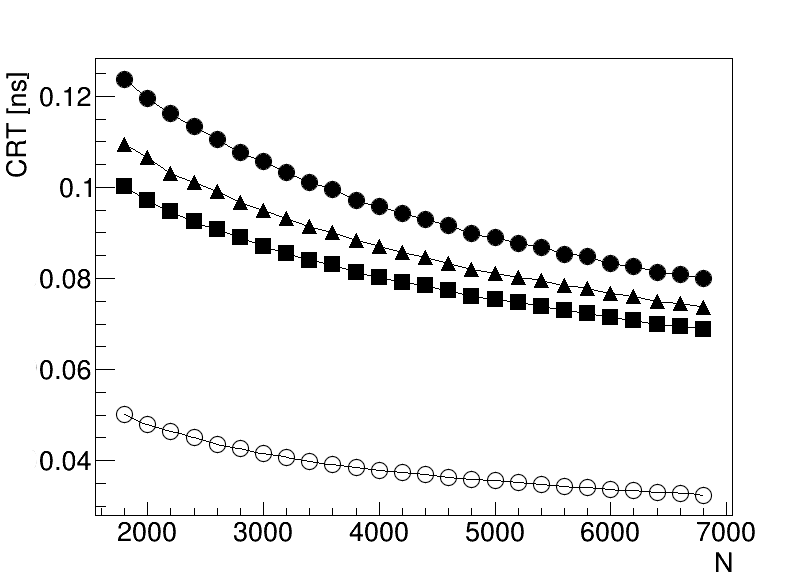}
\end{center}
\caption{
Coincidence resolving time CRT$_{(i)}$
as a function of number of emitted photons N
simulated assuming the emission time distribution of BC-420 plastic scintillator (solid line in Fig.~\ref{emissiontimedistribution}).
Filled points denote results for order statistic $i~=~1$ (circles),
$i~=~2$ (triangles), $i~=~3$ (squares) and 
open circles stands for CRT determined based on the weighted time difference $\sigma(\Delta t)$.
The result does not include the time spread due to the unknown depth-of-interaction.
\label{fig5}
}
\end{figure}

\begin{figure}
\begin{center}
\includegraphics[width=220pt]{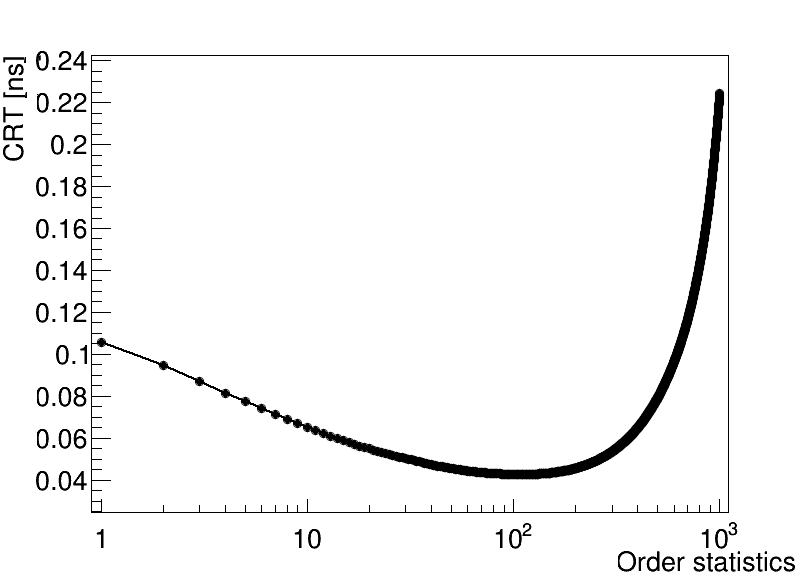}
\end{center}
\caption{
Coincidence resolving time CRT$_{(i)}$
as a function of order statistic 
determined based on simulations of 3000 photons using the 
emission time distribution of BC-420 plastic scintillator (solid line in Fig.~\ref{emissiontimedistribution}). 
The result does not include the time spread due to the unknown depth-of-interaction.
\label{fig6}
}
\end{figure}

In this section we consider 
an ideal plastic scintillator detector where all emitted photons are registered 
by the two ideal photosensors (either left or right) with ideal time resolution 
and 100\% quantum efficiency. It is also assumed that there is no photon absorption and no time spread 
in the infinitely small plastic scintillator. 
In Fig.~\ref{fig5} filled symbols show how 
the coincidence resolving time
for first, second and third order statistic changes as a function of the number of emitted photons,
and Fig.~\ref{fig6} illustrates how 
CRT varies as a function of order statistic in the case of 3000 emitted photons.
As expected from the shape of the probability density distribution of emitted photons (Fig.~\ref{emissiontimedistribution})
the average time difference between emitted photons decreases and hence the time resolution improves 
with the growing order statistic up to the time
when the probability of emission of photons acquires maximum and then time resolution starts to worsen 
since the average time interval between emitted photons increases. 

Having timestamps from all registered photons, in the simplest way we can estimate the hit-time~$\Theta$ 
and time difference~$\Delta t$ e.g. as weighted means
of corresponding values determined for all ordered statistics: 
\begin{equation}
\Theta \equiv \frac{\sum_i \frac{\Theta_{(i)}}{\sigma^2(\Theta_{(i)})}}{\sum_i \frac{1}{\sigma^2(\Theta_{(i)})}}
=\frac{\sum_i \frac{t^L_{(i)}/2}{\sigma^2(\Theta_{(i)})}}{\sum_i \frac{1}{\sigma^2(\Theta_{(i)})}}
+\frac{\sum_i \frac{t^R_{(i)}/2}{\sigma^2(\Theta_{(i)})}}{\sum_i \frac{1}{\sigma^2(\Theta_{(i)})}}
\end{equation}
\begin{equation}
\Delta t \equiv \frac{\sum_i \frac{\Delta t_{(i)}}{\sigma^2(\Delta t_{(i)})}}{\sum_i \frac{1}{\sigma^2(\Delta t_{(i)})}}
=\frac{\sum_i \frac{t^L_{(i)}}{\sigma^2(\Delta t_{(i)})}}{\sum_i \frac{1}{\sigma^2(\Delta t_{(i)})}}
-\frac{\sum_i \frac{t^R_{(i)}}{\sigma^2(\Delta t_{(i)})}}{\sum_i \frac{1}{\sigma^2(\Delta t_{(i)})}}
\end{equation}

Coincidence resolving time CRT$_{(i)}$
is presented in Fig.~\ref{fig5} 
as a function of number of emitted photons 
assuming the probability density distribution for plastic scintillator BC-420 (solid line in Fig.~\ref{emissiontimedistribution}). 

These calculations allow us to find out
what is the best limit of time resolution for the considered detection systems 
when the hit-time is estimated as a weighted mean of the registered timestamps.
Results shown in Fig.~\ref{fig5} imply that in principle 
for the energy deposition in the range from 0.2~MeV (2000 photons) to 0.341~MeV (3410 photons)
a coincidence resolving time is equal to about CRT~=~0.042~ns.

In the following section we will present results of simulations for the solution presently used in the J-PET detector
and compare them with the experimental results. Further on simulations with a matrix SiPM readout will be presented and discussed.

\section{Time resolution for the single module of the J-PET detector}
\begin{figure}
\begin{center}
\includegraphics[width=220pt]{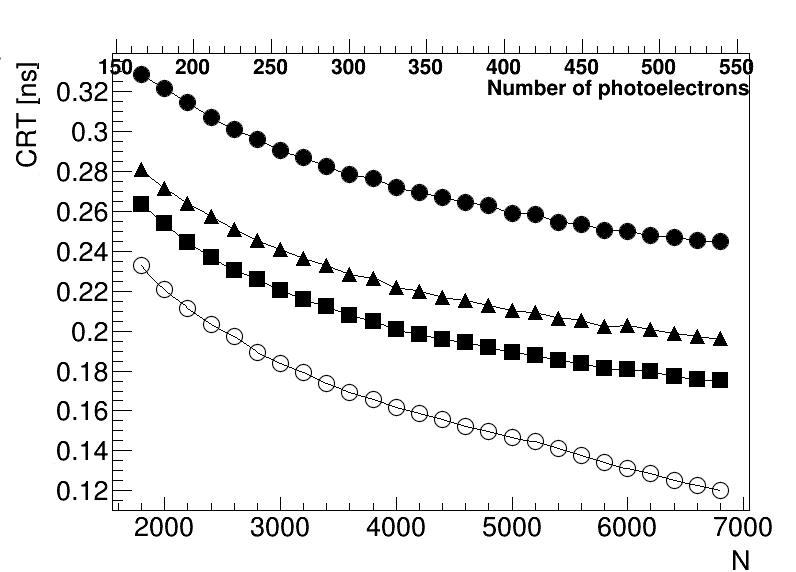}
\end{center}
\caption{
Coincidence resolving time as a function of number of emitted photons N and as a number of photoelectrons 
simulated for the BC-420 plastic scintillator with dimensions of 
0.5~cm~x~1.9~cm~x~30~cm 
read out at two ends by the Hamamatsu R4998 (R5320) photomultipliers.  
Filled points denote results for order statistic $i~=~1$ (circles),
$i~=~2$ (triangles), $i~=~3$ (squares) and 
open circles stand for CRT determined based on the standard deviation of weighted time difference $\sigma(\Delta t)$.
The result does not include the time spread due to the unknown depth-of-interaction.
\label{ModelScin}
}
\end{figure}

\begin{figure}
\begin{center}
\includegraphics[width=220pt]{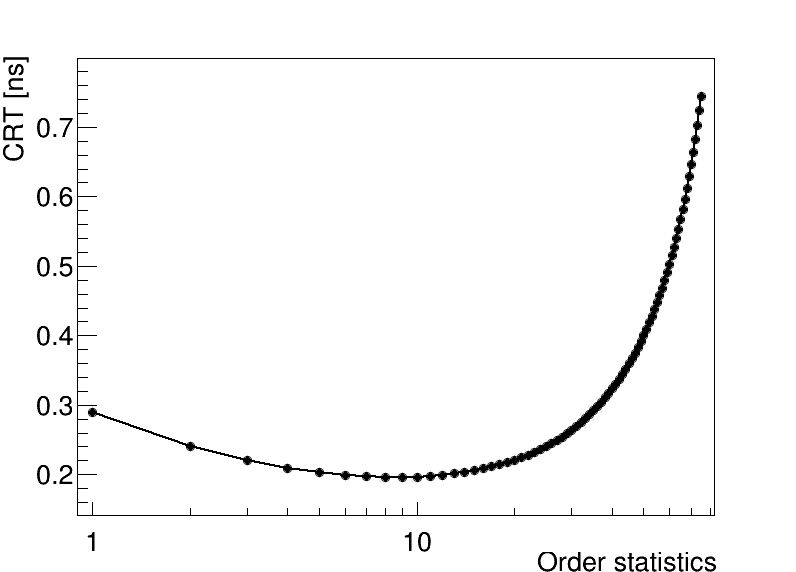}
\end{center}
\caption{
Coincidence resolving time
as a function of order statistic i,
determined based on simulations of 3000 photons using the 
emission time distribution of BC-420 plastic scintillator (solid line in Fig.~\ref{emissiontimedistribution})
with dimensions of 0.5~cm~x~1.9~cm~x~30~cm 
and taking into account a transit time spread of  
the Hamamatsu R4998 (R5320) photomultipliers.  
The result does not include the time spread due to the unknown depth-of-interaction.
\label{ModelScin_3000}
}
\end{figure}

The realistic simulations of the timestamps registered by the large size scintillator detectors 
require a proper account for 
the emission time distribution, photon transport and absorption inside the scintillator, 
as well as quantum efficiency and transit time spread of photosensors.
All these effects have been taken into account as it is described in detail in the Appendix. 
In this section in order to 
test
the simulation procedures
we present results for the plastic scintillator BC-420 
with dimensions of 0.5~cm~x~1.9~cm~x~30~cm 
read out at two ends by the Hamamatsu 
R4998 
(R5320)
photomultipliers. Recent measurements conducted with such detector 
revealed that about 280 photoelectrons are produced from the emission of about 3410 photons corresponding 
to the maximum energy deposition of the 0.511~MeV 
gamma quanta via the Compton effect~
(Moskal et al \hyperlink{Moskal2014}{2014}). 
This is very well reproduced in the simulations as can be inferred from Fig.~\ref{ModelScin} by a comparison 
of values at the upper and lower horizontal axes. Fig.~\ref{ModelScin} shows dependence of the time resolution for the 
first, second and third order statistic as a function of the number of emitted photons. 
The result is consistent with the 
experimental value of 
CRT equal to about 0.266~ns obtained 
when determining time at the threshold -50~mV of the leading edge
of signals corresponding to the range of number of emitted photons between 2000 and 3410~
(Moskal et al \hyperlink{Moskal2014}{2014}).  
Fig.~\ref{ModelScin} indicates that in this range the experimental time resolution of 
0.266~ns is between
the values of time resolutions simulated for the first and third order statistics. 
This is as expected since predominantly only the first few photoelectrons contribute to the 
onset of the leading edge of photomultiplier signals. The obtained result shows that 
in practice the time resolution achievable from the leading edge may be estimated as a mean 
of resolutions for first and third order statistics.
It is also interesting to note that for the discussed detector 
the best time resolution would be obtained by the measurement 
of the tenth ordered statistic (Fig.~\ref{ModelScin_3000}).

\section{Time resolution for plastic scintillator read out by matrices of silicon photomultipliers}
\label{mainsection}
\begin{figure}
\begin{center}
\includegraphics[width=220pt]{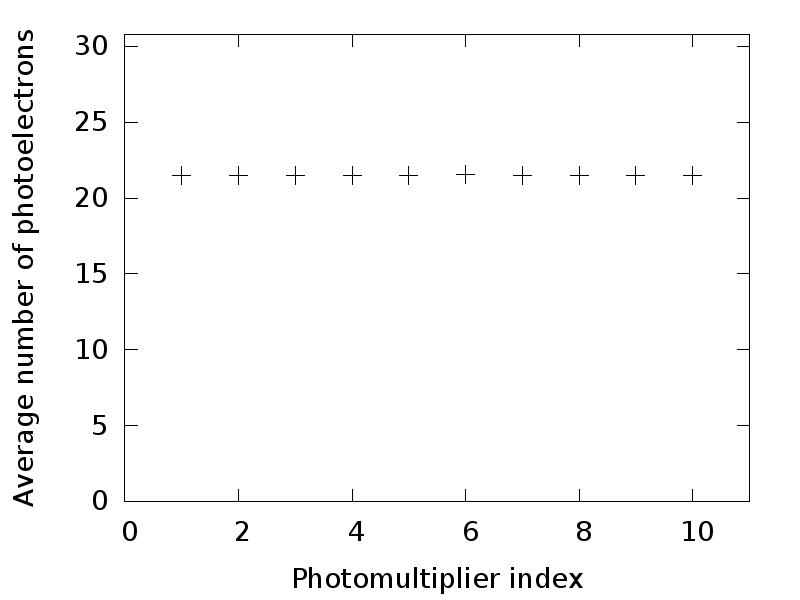}
\end{center}
\caption{ 
Distribution of average number of registered photons as a function of the ID of the photomultiplier.
The simulations were performed for interactions in the center of the scintillator 
with dimensions of 0.7~cm~x~1.9~cm~x~30~cm, 
assuming 3410 photons per interaction, 
corresponding to the maximum energy deposition of 0.511~MeV gamma quanta via Compton effect.
\label{id_distribution}
}
\end{figure}
In the previous sections it was shown that the time resolution may be significantly improved by recording individual timestamps 
of photons arriving to the scintillator edge. 
In this section we present simulation of timing properties achievable
for the long plastic scintillator strips equipped with readouts 
at two sides in the form of a matrix of silicon photomultipliers 
arranged as depicted in Fig.~\ref{detector}. 
The simulations have been performed assuming properties of the Hamamatsu 
S12572-100P silicon photomultiplier~(\hyperlink{Hamamatsu}{Hamamatsu}) 
with photosensitive area of 0.3~cm~x~0.3~cm and the width of non-sensitive rim of 0.05~cm,
and assuming that the scintillator has dimensions of 0.7~cm~x~1.9~cm~x~30~cm. 
A 2~x~5 SIPM matrix (as shown in Fig.~\ref{detector}) 
enables to cover with the photo-sensitive area
about 68\% of the end of scintillator with the cross section of 0.7~cm~x~1.9~cm. 
Such matrices of photomultipliers enable to group photons reaching the end of the scintillator 
into ten subsamples on the left side: 
($t^L_{1,1}, t^L_{1,2}, ..., t^L_{1,N1L}$), ..., ($t^L_{10,1}, t^L_{10,2}, ..., t^L_{10,N10L}$) 
and ten subsamples on the right side:
($t^R_{1,1}, t^R_{1,2}, ..., t^R_{1,N1R}$), ..., ($t^R_{10,1}, t^R_{10,2}, ..., t^R_{10,N10R}$),
where first lower index denotes the ID of SiPM and the second lower index denotes the ID of the photon.
Fig.~\ref{id_distribution} shows that the photons are homogeneously distributed among different photomultipliers
and 
(in the case of the maximum energy deposition by 0.511 MeV gamma quanta)
on the average about 22 photons are registered by each SiPM.

Further on, we assume that a timestamp available from a given SiPM corresponds to the 
time of the fastest photon from the subsample registered by this photomultiplier. 
Therefore, for each subsample separately, we order   
timestamps according to ascending time such that: 
($t^L_{1,(1)} \le t^L_{1,(2)} \le ... \le t^L_{1,(N1L)}$), ...,  ($t^L_{10,(1)} \le t^L_{10,(2)} \le ... \le t^L_{10,(N10L)}$),  
and analogously for the right side,
where indices in brackets indicate timestamps from the ordered subsample.
The fastest timestamps in subsamples: 
$t^L_{1,(1)}, t^L_{2,(1)}, ..., t^L_{10,(1)}$,
and
$t^R_{1,(1)}, t^R_{2,(1)}, ..., t^R_{10,(1)}$
are considered as timestamps registered by the 
photomultipliers (hereafter referred to as photomultiplier's timestamps).
Next, 
for the left and right readout separately, we order the photomultiplier's
timestamps according to ascending time such that: 
($t^L_{[1]} \le t^L_{[2]} \le ... \le t^L_{[10]}$)
and
($t^R_{[1]} \le t^R_{[2]} \le ... \le t^R_{[10]}$),
where indices in square brackets indicate SiPM timestamps after ordering,
and the 
$t_{[i]}$ element in this set will be hereafter referred to as i-th order SiPM statistic. 
For each ordered SiPM statistic the interaction time $\Theta_{[i]}$ 
and the time difference between the signal arrivals 
to the ends of the scintillator $\Delta t_{[i]}$
are estimated as follows:
\begin{equation}
   \Delta t_{[i]} = t^L_{[i]} - t^R_{[i]},
\label{delta_t_id}
\end{equation}
and
\begin{equation}
   \Theta_{[i]} = \frac{t^L_{[i]}+t^R_{[i]}}{2} + const_{[i]},
\label{Theta_i}
\end{equation}
where $const_{[i]}$ will be omitted in the further considerations
without loss of generality. Finally using information available from all 
SiPMs we estimate the
hit-time $\Theta$ and the time difference $\Delta t$ as the weighted mean 
of above defined $\Theta_{[i]}$ and $\Delta t_{[i]}$ values, respectively.

\begin{figure}
\begin{center}
\includegraphics[width=220pt]{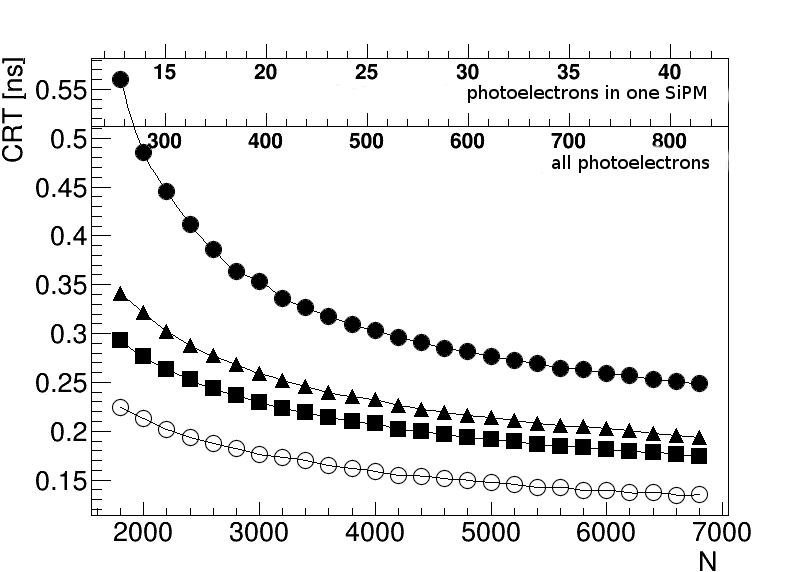}
\end{center}
\caption{
Coincidence resolving time CRT$_{[i]}$
as a function of number of emitted photons N and as a number 
of registered photons (photoelectrons) 
simulated for the BC-420 plastic scintillator with dimensions of 
0.7~cm~x~1.9~cm~x~30~cm 
read out at two ends by a 2~x~5 matrix 
of the Hamamatsu  
S12572-100P silicon photomultipliers.
Filled points denote results for the first, second and third SiPM order statistic: 
$i~=~1$ (circles),
$i~=~2$ (triangles), 
$i~=~3$ (squares) and 
open circles stand for 
CRT determined based on the 
weighted average of all measured $\Delta t_{[i]}$.
\label{silicon_phm}
}
\end{figure}

\begin{figure}
\begin{center}
\includegraphics[width=220pt]{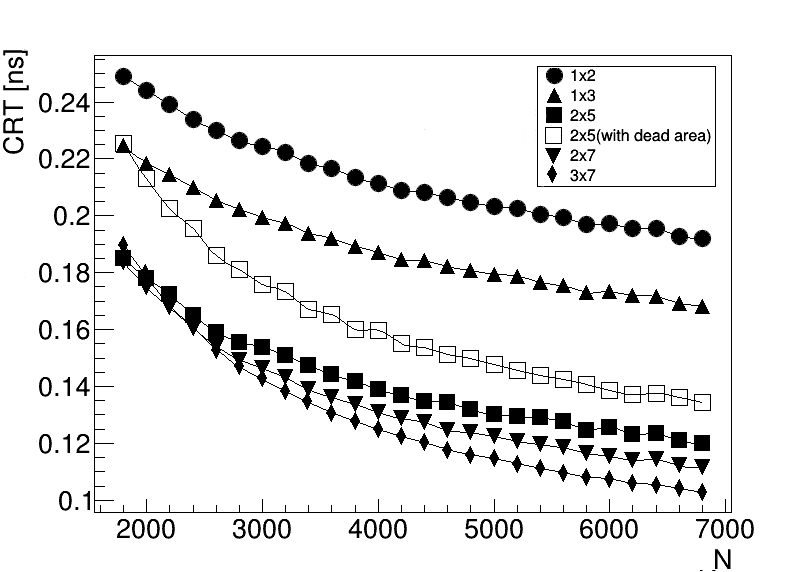}
\end{center}
\caption{
Coincidence resolving time
as a function of number of emitted photons N 
simulated for the BC-420 plastic scintillator with dimensions of 
0.7~cm~x~1.9~cm~x~30~cm. 
In the simulations it was assumed that the readout consists of photomultipliers 
characterized by time spread and quantum efficiency the same as 
the SiPM Hamamatsu  
S12572-100P but with the dimensions allowing to cover fully the 
scintillator with the sensitive area with the following readout configurations:
1~x~2 (filled circles), 
1~x~3 (filled triangles),
2~x~5 (filled squares),
2~x~7 (filled inverted triangles), and 
3~x~7 (filled lozenges).
Open squares indicate results for the 2~x~5 configuration taking into account non-sensitive area 
of the S12572-100P SiPM
(the same as in Fig.~\ref{silicon_phm}).
The result does not include the time spread due to the unknown depth-of-interaction.
\label{silicon_phms}
}
\end{figure}

Results of the performed simulations are shown in Fig.~\ref{silicon_phm}.  
The number of photoelectrons expected for the maximum energy deposition 
of 0.511~MeV gamma quanta (N~=~3410)
is equal to about 440 and is much higher than 280 obtained with the present J-PET prototype.
This increase is due to the higher quantum efficiency of 
S12572-100P silicon photomultipliers with 
respect to the R4998 (R5320) vacuum tube photomultipliers (see Fig.~\ref{fig4} in the Appendix).
Nevertheless, the time resolution for the first SiPM order statistics 
is worse with respect to the one obtainable with the R4998 (R5320) photomultipliers 
because of the larger transit time spread 
of SiPM with respect to R4998 (R5320).
However, due to the access to ten SiPM timestamps (available with the 
2~x~5 SiPM matrix readout),
a coincidence resolving time of CRT~$\approx$~0.180~ns
can be achieved when 
using a weighted mean of the measured SiPM timestamps. This is an average value for the range of interest 
(from 2000 to 3400 photons).
In order to test a dependence of the achievable time resolution on the number of the SiPM in the readout, 
a systematic simulations were conducted changing the number of SiPM from 2 to 21.
Fig.~\ref{silicon_phms} shows 
CRT
obtained  
for various SiPM configurations. 
The result indicates that the improvement of resolution saturates with the growing number of
photomultipliers, and that the 2~x~5 configurations allowing to read 20 timestamps
constitutes an optimal solution and further increase of number of SiPM 
does not improve the resolution significantly. 

\begin{figure}
\begin{center}
\includegraphics[width=220pt]{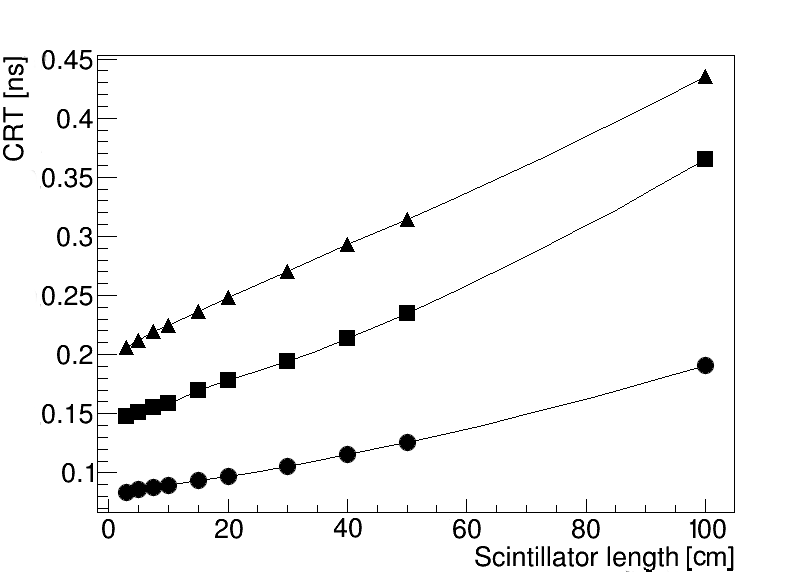}
\end{center}
\caption{
Coincidence resolving time
as a function of the scintillator's length for 2700 emitted photons 
in the center of the scintillator with the cross section of  
0.7~cm~x~1.9~cm.
(Triangles) A mean value of
CRT$_{(1)}$ and 
CRT$_{(3)}$ 
simulated for the scintillator read out at two ends by the Hamamatsu R4998 (R5320) photomultipliers. 
(Squares) 
CRT
simulated for the scintillator read out at two ends by the matrix 
of 2~x~5 S12572-100P photomultipliers.
(Circles) 
CRT 
determined as a weighted mean from all measured timestamps assuming ideal photosensors with no 
transit time spread and 100\% quantum efficiency and assuming that there 
is no photon absorption in the scintillator material.
The shown values take into account an additional smearing of the time due to the unknown depth of interaction. 
This can be well approximated by the FWHM equal to about 0.063~ns in the case of the 1.9~cm thick scintillators. 
The lines are shown to guide the eye.
\label{different_ds}
}
\end{figure}

As a final result in Fig.~\ref{different_ds} a time resolution achievable with 
vacuum photomultipliers R4998 (R5320) is compared to the resolution achievable with 
the 2~x~5 SiPM matrix readout for the length of the scintillators from 2.5~cm up to 100~cm.
Both results are confronted with the resolution 
limit simulated for the ideal photosensors allowing for the measurement 
of each photon reaching the end of the strip.
The result presented in the figure indicates that 
the 2~x~5 SiPM matrix readout can improve the time resolution significantly by about 
a factor of $1.5$ (up to the length of 50~cm) and that still further significant improvement may be 
achieved by increasing the quantum efficiency and decreasing the transit time spread with respect
to the presently available 
S12572-100P silicon photomultiplier produced by ~\hyperlink{Hamamatsu}{Hamamatsu}.
The comparison was done assuming emission of 2700 photons according to the 
spectrum of the EJ-230 (BC-420) scintillator. As it was discussed earlier in the text,
the number of 2700 photons corresponds 
to the average amount of photons useful for the positron emission tomography 
by means of the plastic scintillators.
Finally, the obtained results show that the 2~x~5 S12572-100P matrix readout 
allows to obtain 
CRT~$\approx$~0.366~ns 
even for the J-PET constructed with 
the 100~cm long plastic scintillators. 
%

\section{Summary}
The realistic simulations based on the Monte-Carlo method were conducted 
in order to estimate the time resolution achievable with the J-PET tomography scanner 
built from strips of plastic 
scintillators~(Moskal et al \hyperlink{Moskal2011}{2011},\hyperlink{Moskal2014}{2014},\hyperlink{Moskal2015}{2015},
Raczynski et al \hyperlink{Raczynski2014}{2014},\hyperlink{Raczynski2015}{2015}). 
The simulations took into account: (i) emission spectrum of the plastic scintillator BC-420 (EJ-230),
(ii) probability density distribution of photon emission times, 
(iii) transport of photons along the scintillator strip,
(iv) absorption of photons in the scintillator material,
(v) spectrum of quantum efficiency of photosensors, and (vi) photomultipliers transit time spread.
Arrangement of SiPM photosensors in the form of 2~x~5 matrix attached at two ends to the scintillator strip
allowed for registering 10 timestamps at each side. These after ordering according to the ascending time 
were used to estimate the time of interaction as a weighted mean of times registered for 
each ordered SiPM statistics. Exploitation of information on 10 timestamps at each side improved the time
resolution with respect to the present readout based on vacuum tube photomultipliers
by about a factor of 1.5  despite the fact that the transit time spread of the considered
silicon photosensors S12572-100P 
($\sigma$(TTS)~=~0.128~ns)  
is almost two times larger than TTS
of photomultipliers R4998 (R5320)
($\sigma$(TTS)~=~0.068~ns) 
used in the present version of the J-PET detector.
 
For the energy loss 
in the range from 0.2~MeV to 0.341~MeV  (corresponding to the emission of 2000 to 3410 photons), 
relevant for the positron emission tomography with plastic scintillators,
it was shown that 
with the S12572-100P
photosensors arranged into a 2~x~5 matrix at two ends of the scintillator strip
the coincidence resolving time changes from CRT~$\approx$~0.170~ns to CRT~$\approx$~0.365~ns
when extending an axial field-of-view from
15~cm to 100~cm. 
This corresponds to the changes of the axial position resolution from 1.4~cm (FWHM)  
to 3.1~cm (FWHM), respectively. However, as it is shown by solid circles in Fig.~\ref{different_ds} 
there is still room for improving  CRT and  hence also for improving an axial position resolution 
by about a factor of two by decreasing the time-jitter of the SiPMs. 
The results open perspectives for construction of the cost-effective TOF-PET scanner
with significantly better TOF resolution and larger field-of-view  
with respect to the newest TOF-PET modalities
characterized by 
CRT~$\approx$~0.4~ns~(\hyperlink{Philips}{Philips},\hyperlink{GeneralElectric}{General Electric}).
%
In addition, a J-PET scanner built from long strips of plastic scintillators 
read out by the silicon photosensors, may be combined with the Magnetic Resonance Imaging modality
in a way allowing for the simultaneous PET and MRI measurement 
with the large field-of-view~(Moskal \hyperlink{Moskal2014b}{2014b}) 
e.g. by inserting a barrel built of plastic strips into the MRI system. 

Finally, it was shown that  not only an intrinsic lower bound for the time resolution
calculated 
using 
the Fisher information and 
Cram\'e{}r-Rao
inequality, but also more practical limit determined for the 
time estimated as a mean of all timestamps registered with the ideal photosensor
is much lower than the above quoted resolutions.
Therefore, there is still room for further improvement of the TOF resolution 
of the J-PET tomograph
which can be achieved anticipating
future availability of silicon photosensors with transit-time-spread lower than $\sigma$(TTS)~=~0.128~ns
of S12572-100P Hamamatsu photomultipliers.

The main purpose of the development of the J-PET system is to find a cost-effective 
way of the whole body PET imaging. Thus, in order to compare the performance of the J-PET with 
the presently available LSO based TOF-PET devices, by analogy 
to references~(Conti \hyperlink{Conti2009}{2009}, Eriksson Conti \hyperlink{ErikssonConti2015}{2015}) 
we introduce a following formula expressing a figure-of-merit $FOM_{wb}$
relevant for the whole body imaging:
\begin{equation}
FOM_{wb} ~= \epsilon_{detection}^2 \times \epsilon_{selection}^2 \times Acc \ / \ (CRT \times N_{steps}),
\label{FOM}
\end{equation}
where
$\epsilon_{detection}$ denotes the detection efficiency of a single 0.511 MeV gamma quantum,

$\epsilon_{selection}$ indicates the selection efficiency of “image-forming” events,

$CRT$  denotes the coincidence resolving time,

$N_{steps}$ indicates number of steps (bed positions) needed to scan a whole body, and 

$Acc$ denotes a geometrical acceptance.

In the first order of approximation we may assume that
$N_{steps}$ is inversely proportional to the AFOV,   
and that 
the term 
$\epsilon_{detection}^2 \times Acc$
is proportional to the 
$\int_{\theta_{min}}^{\theta_{max}} (\epsilon_{0detection}/\sin\!\theta)^2 \sin\!\theta d\theta$,
where $\theta$ denotes the angle between the direction of the gamma quanta and the main
axis of the tomograph;  the term $\epsilon_{0detection}/\sin\!\theta$
accounts for the changes of the detection efficiency as a function of the $\theta$ angle,
with 
$\epsilon_{0detection}$ denoting detection efficiency when gamma quantum crosses 
the detector perpendicularly to its surface;  
$\sin\!\theta d\theta$
stands for the angular dependence of the differential element of the solid angle,
and
$\theta_{min}$ to $\theta_{max}$
determines the range of angular acceptance of the tomograph.
The above assumptions yield: 
\begin{equation}
FOM_{wb} ~= 
\int_{\theta_{min}}^{\theta_{max}} (\epsilon_{0detection}/\sin\!\theta)^2 \sin\!\theta d\theta
\times AFOV \ / \ CRT.
\label{FOM2}
\end{equation}
\begin{figure}[h]
\begin{center}
\includegraphics[width=250pt]{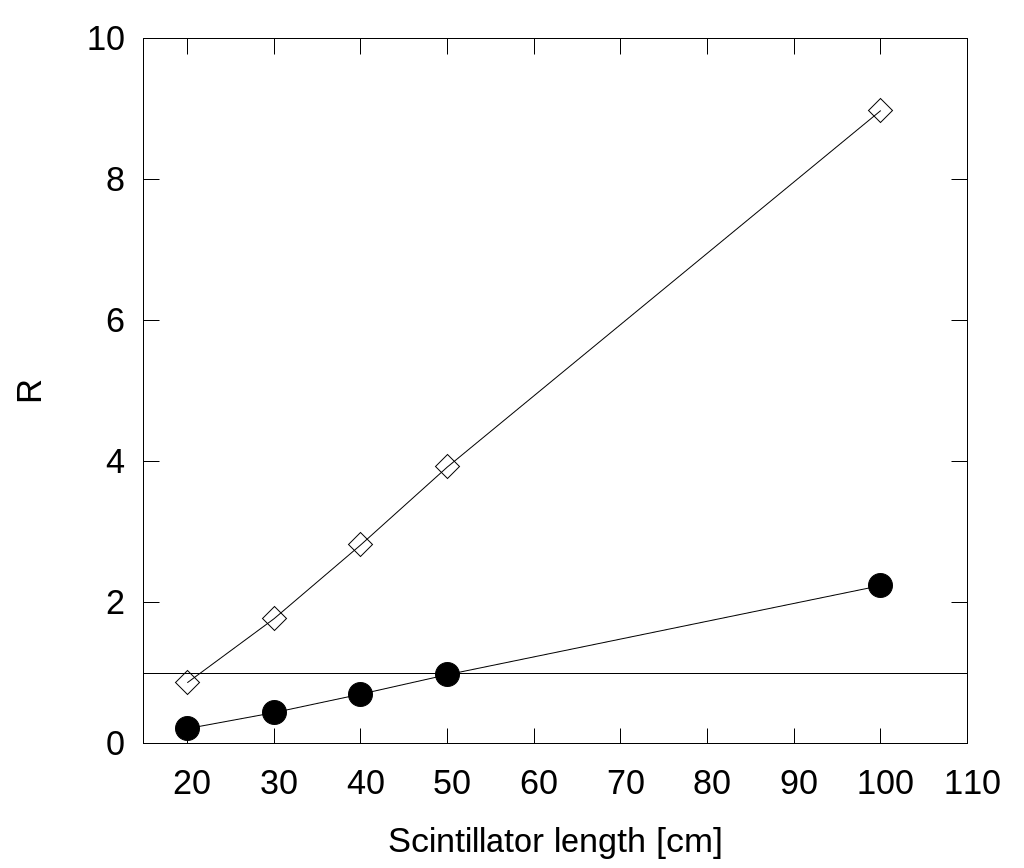}
\end{center}
\caption{A ratio of figure of merits for the whole body imaging with J-PET and LSO based PET detectors defined as  
R(AFOV-J-PET) = $FOM_{wb}$(AFOV-J-PET)~/~$FOM_{wb}$(LSO with AFOV=20cm). 
Horizontal axis of the figure refers to the length of the J-PET detector. The length of the LSO scanner was fixed to 20~cm,
and the diameter of the scanners was fixed to 80~cm.
Full dots indicate result determined  for a single J-PET layer  ($N_{layers}$ = 1),  
and  open squares indicate result for J-PET with two layers ($N_{layers}$ = 2).  
For small number of layers, the $\epsilon_{detection}$ of the J-PET detector 
is approximately proportional to $N_{layers}$. The presented result was obtained under assumptions
that CRT of the LSO based detectors is equal to 0.4 ns and that CRT values of the J-PET corresponds to the
results of this article shown by full squares in Fig.~\ref{different_ds}.
Calculations of $FOM_{wb}$ of the J-PET were performed assuming a threshold of 0.2~MeV.
The lines connecting points are shown to guide the eye,
whereas the horizontal solid line indicates R~=~1.
\label{R-FOM}
}
\end{figure}
Fig.~\ref{R-FOM} shows the ratio $R$ of $FOM_{wb}$ of the J-PET and LSO based PET detectors defined as:
R(AFOV-J-PET) = $FOM_{wb}$(AFOV-J-PET)/$FOM_{wb}$(LSO with AFOV~=~20~cm). 
The shown ratio is determined for a fixed AFOV~=~20~cm of the LSO based PET, 
but  varying the AFOV of the J-PET detector.   
Full dots indicate result determined  for a single J-PET layer  ($N_{layers}$~=~1),  
and  open squares indicate result for J-PET with two layers ($N_{layers}$~=~2).  
The results shown in Fig.~\ref{R-FOM} were obtained assuming a 2~cm radial thickness of the detection 
layers and  taking into account that linear attenuation coefficients of 0.511 MeV gamma quanta 
are equal to $\mu_{LSO}~=~0.87~cm^{-1}$~(Mechler \hyperlink{Mechler2000}{2000}) 
and  $\mu_{plastic}~=~0.098~cm^{-1}$~(\hyperlink{SaintGobain}{Saint Gobain Crystals}). 
Furthermore it was assumed that (i) CRT of LSO based scanners is equal to 0.4~ns 
as achieved recently by manufacturers Philips~(\hyperlink{Philips}{Philips}) 
and General Electric~(\hyperlink{GeneralElectric}{General Electric}), 
and that (ii) $\epsilon_{selection}$ for LSO is equal to 0.32,  
which is a fraction of the photoelectric effect in the case of  the LSO crystals [Humm2003], 
and that (iii) $\epsilon_{selection}$ of the J-PET is equal to 0.44  which corresponds to the fraction 
of events with energy deposition larger  than 0.2 MeV in the case of plastic scintillators.  
Fig.~\ref{R-FOM} indicates that in order to compensate for the lower efficiency 
of plastic scintillators and thus to obtain $FOM_{wb}$ of the J-PET comparable to the LSO based scanners 
with AFOV~=~20~cm it is required to use either two detection layers or to increase the J-PET AFOV to about 50~cm.
Certainly the  $FOM_{wb}$ of the LSO based PET would also grow approximately as square of AFOV but at the same time 
the cost of such PET detector would increase almost linearly proportional to AFOV, 
whereas the cost of the J-PET does not increase significantly when increasing the AFOV.  

The relative ease of the cost effective increase of the axial field-of-view 
makes the J-PET tomograph competitive with respect to the current commercial PET scanners 
as regards sensitivity and time resolution, 
yet this is achieved at the expense of the significant reduction of the axial spatial resolution.  
Finally, it is worth to stress that the J-PET with a long diagnostic chamber opens unique 
perspectives for simultaneous whole-body metabolic imaging not accessible with the presently 
available PET/CT modalities.

\section{Acknowledgements}
We acknowledge technical and administrative support of T.~Gucwa-Rys,
A.~Heczko, M.~Kajetanowicz, G.~Konopka-Cupia{\l},  W.~Migda{\l},
and the financial support by The Polish National Center for 
Research and Development through grant INNOTECH-K1/IN1/64/159174/NCBR/12,
the Foundation for Polish Science through MPD programme, the EU and MSHE
Grant No. POIG.02.03.00-161 00-013/09, Doctus - the Lesser Poland PhD
Scholarship Fund, and Marian Smoluchowski Krakow Research Consortium
"Matter-Energy-Future". 

\section{References}
\leftskip= 1cm
\parindent= -1cm
\par
\hypertarget{3Mcom}
        3M Optical Systems, 
        \textit{www.3M.com/Vikuiti}
\par
\hypertarget{Baszak2014} 
         Baszak~J
         2014
         Hamamatsu, private communication.
\par
\hypertarget{Bettinardi2011} 
        Bettinardi~V et al
        2011 
        Physical Performance of the new hybrid PET-CT Discovery-690 FWH-TOF=544ps
        \textit{Med. Phys.} {\bf 38} 5394-5411
\par
\hypertarget{Conti2009} 
         Conti~M 
         2009
         State of the art and challenges of time-of-flight PET
         \textit{Phys. Med.} {\bf 25} 1-11
\par
\hypertarget{Conti2011}
        Conti~M 
        2011
        Focus on time-of-flight PET: the benefits of improved time resolution
        \textit{Eur. J. Nucl. Med. Mol. Imaging} {\bf 38} 1147-1157
\par
\hypertarget{ContiEriksson2009} 
		Conti~M, Eriksson~L, Rothfuss~H, Melcher~C~L
		2009
        Comparison of Fast Scintillators With TOF PET Potential  
        \textit{IEEE Trans. Nucl. Sci.} {\bf 56} 926-933
\par
\hypertarget{DeGroot1986}
       DeGroot~M~H
       1986
       \textit{Probability and Statistics Addison-Weslay 420-6}
\par
\hypertarget{Eljen} 
        Eljen Technology, 
        \textit{www.eljentechnology.com}
\par
\hypertarget{ErikssonConti2015} 
		Eriksson~L and Conti~M 
		2015
        Randoms and TOF gain revisited
        \textit{Phys. Med. Biol.} {\bf 60} 1613-1623
\par
\hypertarget{Fishburn2010}
         Fishburn~M~W and Charbon~E
         2010
         System Tradeoffs in Gamma-Ray Detection Utilizing SPAD Arrays and Scintillators
         \textit{IEEE Trans. Nucl. Sci.} {\bf 57} 2549-2557
\par
\hypertarget{GeneralElectric} 
         General Electric,
         \textit{http://www3.gehealthcare.com/en/products/categories/magnetic\_resonance\_imaging/signa\_pet-mr}
\par
\hypertarget{Hamamatsu} 
        Hamamatsu, 
        \textit{www.hamamatsu.com}
\par
\hypertarget{Humm2003} 
         Humm~J~L, Rosenfeld~A, Del~Guerra~A
         2003
         From PET detectors to PET scanners
         \textit{Eur.  J.  Nucl.  Med.  Mol.  Imaging} {\bf 30} 1574-1597
\par
\hypertarget{Karp2008}
         Karp~J~S et al
         2008
         Benefit of Time-of-Flight in PET: Experimental and Clinical Results  
         \textit{J.  Nucl.  Med.} {\bf 49} 462-470
\par
\hypertarget{Korcyl2014}
        Korcyl~G et al
        2014
        Trigger-less and reconfigurable data acquisition system for positron emission tomography
        \textit{Bio-Algorithms and Med-Systems} {\bf 10} 37-40
\par
\hypertarget{Kowalski2015}
         Kowalski~P et al
         2015
         Multiple scattering and accidental coincidences in the J-PET detector simulated using GATE package
         \textit{Acta Phys. Pol.} {A 127} 1505-1512 [arXiv:1502.04532 [physics.ins-det]].
\par
\hypertarget{Kuhn2006}
		Kuhn~A et al
		2006
        Performance assessment of pixelated LaBr3 detector modules for time-of-flight PET 
        \textit{IEEE Trans. Nucl. Sci.} {\bf 53} 1090-1095
\par
\hypertarget{Lecoq2013}
         Lecoq~P, Auffray~E, Knapitsch~A
         2013
         How Photonic Crystals Can Improve the Timing Resolution of Scintillators
         \textit{IEEE Trans. Nucl. Sci.} {\bf 60} 1653-1657
\par
\hypertarget{Mechler2000}
         Mechler~C~L
         2000
         Scintillation Crystals for PET,
         \textit{J. Nucl. Med.} {\bf 41} 1051-1055
\par
\hypertarget{Meijlink2011}
        Meijlink~J~R et al
        2011
        First Measurement of Scintillation Photon Arrival Statistics Usign a High-Granularity Solid-State 
        Photosensor Enabling Time-Stamping of up to 20,480 Single Photons
        \textit{Proc. IEEE Nuclear Science Symposium (NSS/MIC).}  2254-2257
\par
\hypertarget{Moses1999} 
		Moses~W~W, Derenzo~S~E
		1999
        Prospects for Time-of-Flight PET using LSO Scintillator
        \textit{IEEE Trans. Nucl. Sci.} {\bf 46} 474-478
\par
\hypertarget{Moses2003}
		Moses~W~W
		2003
        Time of Flight in PET Revisited 
        \textit{IEEE Trans. Nucl. Sci.} {\bf 50} 1325-1330
\par
\hypertarget{Moskal2011}
        Moskal~P et al
        2011
        Novel detector systems for the Positron Emission Tomoraphy
        \textit{Bio-Algorithms and Med-Systems} {\bf 7} 73-78; [arXiv:1305.5187 [physics.med-ph]].
\par
\hypertarget{Moskal2012}
       Moskal~P et al
       2012
       TOF-PET detector concept based on organic scintillators
       \textit{Nuclear Medicine Review} {\bf 15} C81-C84; [arXiv:1305.5559 [physics.ins-det]].
\par
\hypertarget{Moskal2014}
         Moskal~P et al
         2014
         Test of a single module of the J-PET scanner based on plastic scintillators 
         \textit{Nucl. Instrum. Methods Phys. Res.} {\bf A 764} 317-321; [arXiv:1407.7395 [physics.ins-det]].
\par
\hypertarget{Moskal2014b}
         Moskal~P
         2014b 
         A hybrid TOF-PET/MRI tomograph 
         \textit{Patent Application} PCT/EP2014/068373 WO2015028603 A1
\par
\hypertarget{Moskal2015}
         Moskal~P et al
         2015
         A novel method for the line-of-response and time-of-flight reconstruction 
         in TOF-PET detectors based on a library of synchronized model signals 
         \textit{Nucl. Instrum. Methods Phys. Res.} {\bf A 775} 54-62; [arXiv:1412.6963 [physics.ins-det]].
\par
\hypertarget{Moszynski1977}
        Moszynski~M and Bengtson~B
        1977
        Light pulse shapes from plastic scintillators
        \textit{Nucl. Instrum. Methods} {\bf 142} 417-434
\par
\hypertarget{Moszynski1979}
        Moszynski~M and Bengtson~B
        1979
        Status of timing with plastic scintillation detectors 
        \textit{Nucl. Instrum.  Methods} {\bf 158} 1-31
\par
\hypertarget{Moszynski2011} 
          Moszynski~M, Szczesniak~T
          2011
          Optimization of Detectors for Time-of-Flight PET  
          \textit{Acta Phys. Pol. B. Proc. Supp.} {\bf 4} 59-64
\par
\hypertarget{Nickles1978}
          Nickles~R~J, Meyer~H~O
          1978
          Design of a three-dimensional positron camera for nuclear medicine
          \textit{Phys. Med. Biol.} {\bf 23} 686-695 
\par
\hypertarget{Palka2014}
        Palka~M et al
        2014
        A novel method based solely on FPGA units enabling measurement of time and charge 
        of analog signals in Positron Emission Tomography  
        \textit{Bio-Algorithms and Med-Systems} {\bf 10} 41-45; [arXiv:1311.6127 [physics.ins-det]].
\par
\hypertarget{Philips} 
         Philips,
         \textit{http://www.philips.co.uk/healthcare/product/HC882446/vereos-digital-pet-ct}
\par
\hypertarget{Raczynski2014}
         Raczy{\'n}ski~L et al
         2014
         Novel method for hit-position reconstruction using voltage signals in plastic scintillators 
         and its application to Positron Emission Tomography 
         \textit{Nucl. Instrum. Methods Phys. Res.} {\bf A 764} 186-192; [arXiv:1407.8293 [physics.ins-det]].
\par
\hypertarget{Raczynski2015}
         Raczy{\'n}ski~L et al
         2015
         Compressive Sensing of Signals Generated in Plastic Scintillators in a Novel J-PET Instrument 
         \textit{Nucl. Instrum. Methods Phys. Res.} {\bf A 786} 105-112; [arXiv:1503.05188 [physics.ins-det]].
\par
\hypertarget{SaintGobain}
        Saint Gobain Crystals 
        \textit{www.crystals.saint-gobain.com}
\par
\hypertarget{Seifert2012}
       Seifert~S, van~Dam~H~T, Schaart~D~R
       2012
       The lower bound on the timing resolution of scintillation detectors
       \textit{Phys. Med. Biol.} {\bf 57} 1797-1814
\par
\hypertarget{Seifert2012b}
         Seifert~S,van~Dam~H~T, Vinke~R, Dendooven~P, Lohner~H, Beekman~F~J, Schaart~D~R
         2012
         A Comprehensive Model to Predict the Timing Resolution of SiPM-Based Scintillation Detectors: 
         Theory and Experimental Validation
         \textit{IEEE Trans. Nucl. Sci.} {\bf 59} 190-204
\par
\hypertarget{Schaart2009}
		Schaart~R~D et al
		2009
        A novel, SiPM-array-based, monolithic scintillator detector for PET 
        \textit{Phys. Med. Biol.} {\bf 54} 3501-3512
\par
\hypertarget{Schaart2010} 
		Schaart~R~D et al
		2010
        LaBr3:Ce and SiPMs for TOF PET: achieving 100ps coincidence resolving time.
        \textit{Phys.  Med.  Biol.} {\bf 55} N179-N189
\par
\hypertarget{Senchyshyn2006}
         Senchyshyn~V et al
         2006
         Accounting for self-absorption in calculation of light collection
         in plastic scintillators
         \textit{Nucl. Instrum. Methods Phys. Res.} {\bf A 566} 286-293
\par
\hypertarget{Spanoudaki2011}
        Spanoudaki~V~Ch and Levin~C~S
        2011
        Investigating the temporal resolution limits of scintillation 
        detection from pixellated elements: comparison between experiment and simulations
        \textit{Phys. Med. Biol.} {\bf 56} 735-756
\par
\hypertarget{Surti2007} 
                  Surti~S et al 
                  2007
         Performance of Philips Gemini TF PET/CT Scanner with Special Consideration 
         for Its Time-of-Flight Imaging Capabilities
         \textit{J. Nucl. Med.} {\bf 48} 471-480 
\par
\hypertarget{Szymanski2014} 
      Szyma{\'n}ski~K et al
      2014
      Simulations of gamma quanta scattering in a single module of the J-PET detector 
      \textit{Bio-Algorithms and Med-Systems} {\bf 10} 71-77
\par
\hypertarget{Townsend2004} 
		 Townsend~D~W
		 2004
         Physical Principles and Technology of Clinical PET Imaging
         \textit{Ann. Acad. Med. Singapore} {\bf 33} 133-145
\par

\section*{APPENDIX: Simulation of photon transport in cuboidal scintillator strips}
In a long scintillator strip, a photon on its way from the emission point to the photomultiplier
may undergo many internal reflections whose number strongly depends on the scintillator size and the photon emission angle. 
However, the space reflection symmetries of the cuboidal shapes, which are considered in this article,
enables a significant simplification of the photon transport algorithm, without 
following photon propagation in a typical manner.
In our simulations for each emitted photon
the initial direction of its flight is obtained in polar coordinate 
system as two uniformly distributed random values of $cos\theta$ and $\phi$, 
where $\theta$ is the angle between flight direction and $z$-axis
and $\phi$ is the azimuthal angle as defined in standard spherical coordinate system.
The coordinate system is defined in Fig.~\ref{detector} where the $z$-axis is directed along the longest axis of the scintillator strip.
The components of photon flight direction vector can be expressed as follows:
\begin{equation}
\vec{dir} = [sin\theta cos\phi,~sin\theta sin\phi,~cos\theta]
\end{equation}
The number of reflections from the side surfaces that are normal to $x$ or $y$-axis 
are calculated using the projection of the flight-direction vector to $y$-$z$ or $x$-$z$-plane, respectively:
\begin{equation}
tg\theta_x=\frac{dir_x}{dir_z};~tg\theta_y=\frac{dir_y}{dir_z},
\end{equation}
{\noindent where $\theta_x$ is the angle between $\vec{dir}$ projection on $x$-$z$-plane and $z$-axis and $\theta_y$ is the same for projection on $y$-$z$-plane.}\\

\begin{figure}
\begin{center}
\includegraphics[width=220pt]{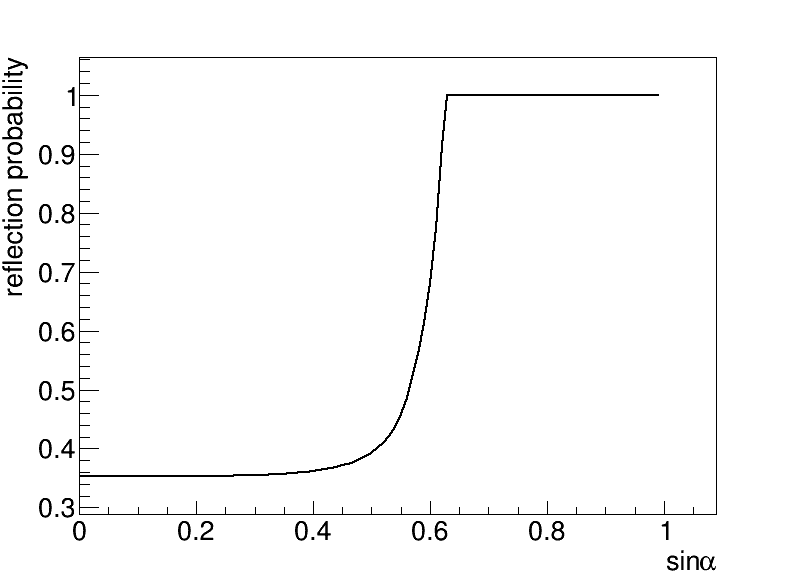}
\end{center}
\caption{
The dependence of reflection probability on sinus of the reflection angle $\alpha$.
\label{fig2}
}
\end{figure}

\begin{figure}
\begin{center}
\includegraphics[width=220pt]{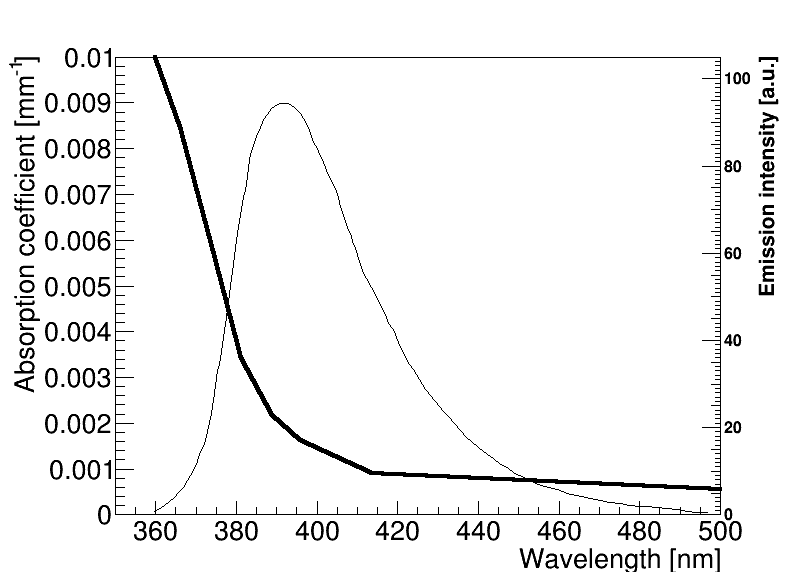}
\end{center}
\caption{
(Thick line) The dependence of scintillator's absorption coefficient $\mu_{eff}$ on photon's wavelength.
The effective coefficient $\mu_{eff}$ was determined by scaling 
the absorption coefficient of the pure 
polystyrene~(Senchyshyn et al 2006) by factor of 1.8~(Kowalski et al 2015).
(Thin line) Emission spectrum of the BC-420 plastic scintillator~(\protect\hyperlink{SaintGobain}{Saint Gobain Crystals}).
The left axis denotes absorption coefficient and right axis denotes the emission intensity. 
\label{fig3}
}
\end{figure}

Taking into account the fact that each reflection changes only the sign of respective component of $\vec{dir}$ 
we can assume that the reflection angle is not changed for each pair of side surfaces during the whole photon flight.
So we need to obtain two values of reflection angle:
one for side surfaces normal to $x$-axis and one for ones normal to $y$-axis.
Knowing that $|\vec{dir}|=1$ and that these are the angles between photon 
flight direction and the normal vectors for respective side surfaces ($x$ and $y$ axes) we obtain:
\begin{equation}
cos\alpha_x=dir_x;~cos\alpha_y=dir_y,
\end{equation}
where $\alpha_{x}$ and
 $\alpha_{y}$ 
  are the reflection angles for side surfaces normal to respective axis.
Then the probability of photon's reaching the photomultiplier can be calculated using a following formula:
\begin{equation}
P_{reach}=P_{refl}(sin\alpha_x)^{n_x} P_{refl}(sin\alpha_y)^{n_y},
\label{flightformula}
\end{equation}
where $n_{x}$ 
and
$n_{y}$ 
denote the respective numbers of reflections.
The dependence $P_{refl}(sin\alpha)$ is obtained from Fresnel equations
and is shown in Fig.~\ref{fig2}.
Further factors that influence the photon registration probability 
are absorption in the scintillator material, losses at surface imperfections,
 and the photomultiplier's 
quantum efficiency.
In current algorithm of simulation the following formula for photon registration probability has been used
\begin{equation}
P_{reg} = P_{reach}~\epsilon(\lambda)~e^{-{\mu_{eff}(\lambda)}~\frac{\Delta L}{cos \theta}}
\end{equation}
where $P_{reach}$ denotes the probability from formula (\ref{flightformula}), 
$\lambda$ denotes the photon's wavelength,
$\epsilon(\lambda)$ stands for the photomultiplier's quantum efficiency and $\mu_{eff}(\lambda)$ 
is the effective absorption coefficient for the scintillator material.
The latter, shown by thick solid line in Fig.~\ref{fig3}, accounts effectively for the absorption 
of photons on the way to photomultipliers 
and was determined by scaling the 
absorption coefficient of pure polystyrene~(Senchyshyn et al \hyperlink{Senchyshyn2006}{2006})
to the experimental results obtained with the single detection unit of the J-PET detector~
(Kowalski et al \hyperlink{Kowalski2015}{2015}).
The scaling factor accounts effectively for the absorption due
to the primary and secondary admixture in the scintillator material, 
imperfections of surfaces and reflectivity of the foil~(Kowalski et al \hyperlink{Kowalski2015}{2015}).
It was determined by the comparison of simulations with experimental results obtained 
for the EJ-230 plastic scintillator with dimensions 
of 0.5~cm~x~1.9~cm~x~30~cm~(Kowalski et al \hyperlink{Kowalski2015}{2015}). 

\begin{figure}
\begin{center}
\includegraphics[width=220pt]{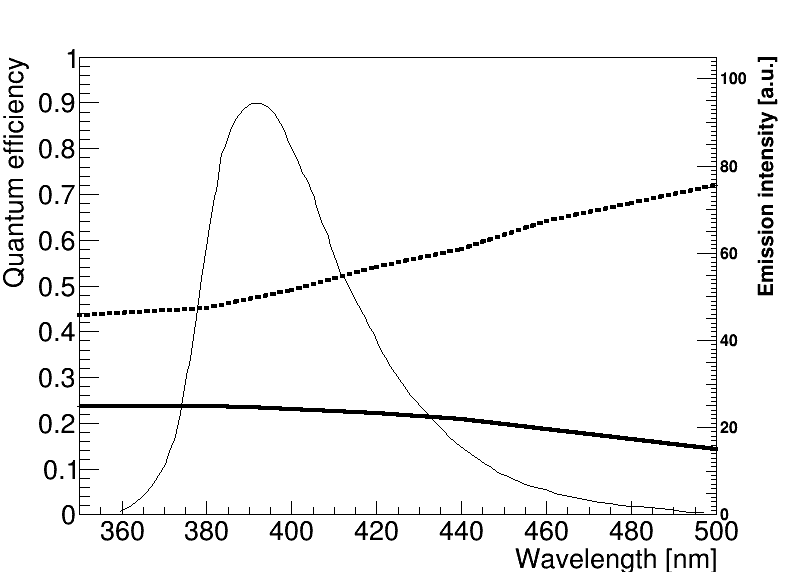}
\end{center}
\caption{
Quantum efficiency as a function of photon's wavelength for Hamamatsu R4998 (R5320) photomultiplier (Baszak 2014)
(solid line)
and Hamamatsu silicon 
S12572-100P photomultiplier (dashed line).
A superimposed thin solid line denotes emission spectrum 
of the Saint Gobain BC-420 plastic scintillator.
The left axis denotes quantum efficiency and right axis denotes the emission intensity. 
\label{fig4}
}
\end{figure}

Photomultipliers quantum efficiencies that were used in current calculations are shown in 
Fig.~\ref{fig4}.

The time 
of arrival 
$t_i^{arrival}$ 
of $i$th photon at the photomultiplier 
(or in general the time of passing a given distance ${\Delta L}$ along the scintillator)
may be expressed as:
\begin{equation}
t_i^{arrival} = t_i^{e} + \frac{\Delta L}{\frac{c}{n} cos \theta},
\label{time_phm}
\end{equation}
where $t_i^{e}$ is the emission time of $i$th photon,
$\Delta L$ denotes the distance between the emission point and the photomultiplier, 
$c$ denotes the speed of light 
and $n$ stands for scintillator's refractive index (the value of $n = 1.58$ was used)
(\hyperlink{SaintGobain}{Saint Gobain Crystals}).

Finally, the timestamp $t_i$ is simulated by smearing the time 
$t_i^{arrival}$ 
taking into account the transition time spread of the photosensors:
\begin{equation}
t_{i} = t_i^{arrival} + RG(t_{offset}, \sigma_t),
\end{equation}
where  $RG(0,\sigma)$ is value generated randomly according to the Gauss distribution 
with the mean at $t_{offset}$, 
and with standard deviation $\sigma_t$ 
equal to the standard deviation of time spread of a given photomultiplier.
For simulations referred to in this paper as done with "ideal photomultiplier" $\sigma_t = 0$.
Otherwise for Hamamatsu R4998 (R5320) photomultiplier $\sigma_t = 0.068~ns$~\hyperlink{Hamamatsu}{Hamamatsu}
and for silicon photomultiplier S12572-100P $\sigma = 0.128~ns$~\hyperlink{Hamamatsu}{Hamamatsu}.
The parameter $t_{offset}$ accounts for all constant electronic time delays and 
its value does not influence the time resolution.
Therefore, for simplicity, 
but without loss of generality, it is set to zero.

\end{document}